\documentclass{emulateapj}
\DeclareGraphicsExtensions{.pdf,.png,.jpg}
\usepackage{amsmath}

\shorttitle{M51, NGC891}
\shortauthors{Jones et al.}

\begin{document}

\title{HAWC+ Far-Infrared Observations of the Magnetic Field Geometry in M51 and NGC 891.}

\author{Terry Jay Jones}
\affil{Minnesota Institute for Astrophysics, University of Minnesota, Minneapolis, MN 55455, USA}

\author{Jin-Ah Kim}
\affil{Minnesota Institute for Astrophysics, University of Minnesota, Minneapolis, MN 55455, USA}

\author{C. Darren Dowell}
\affil{NASA Jet Propulsion Laboratory, California Institute of Technology, 4800 Oak Grove Drive, Pasadena, CA 91109, USA}

\author{Mark R. Morris}
\affil{Department of Physics and Astronomy, University of California, Los Angeles, Box 951547, Los Angeles, CA 90095-1547 USA}

\author{Jorge L. Pineda}
\affil{Jet Propulsion Laboratory, California Institute of Technology, 4800 Oak Grove Drive, Pasadena, CA 91109, USA}

\author{Dominic J. Benford}
\affil{NASA Headquarters, 300 E Street SW, Washington DC  20546, USA}

\author{Marc Berthoud}
\affil{Engineering + Technical Support Group, University of Chicago, Chicago, IL 60637, USA}
\affil{Center for Interdisciplinary Exploration and Research in Astrophysics (CIERA), and Department of Physics \& Astronomy, Northwestern University, 2145 Sheridan Rd, Evanston, IL, 60208, USA}

\author{David T. Chuss}
\affil{Department of Physics, Villanova University, 800 E. Lancaster Ave., Villanova, PA 19085, USA}

\author{Daniel A. Dale}
\affil{Department of Physics \& Astronomy, University of Wyoming, Laramie, WY, USA}

\author{L. M. Fissel}
\affil{Department of Physics, Engineering Physics and Astronomy, Queenx92s University, 64 Bader Lane, Kingston, ON, Canada, K7L 3N6}

\author{Paul F. Goldsmith}
\affil{NASA Jet Propulsion Laboratory, California Institute of Technology, 4800 Oak Grove Drive, Pasadena, CA 91109, USA}

\author{Ryan T. Hamilton}
\affil{Lowell Observatory, 1400 W Mars Hill Rd, Flagstaff, AZ 86001, USA}

\author{Shaul Hanany}
\affil{School of Physics and Astronomy, University of Minnesota / Twin Cities, Minneapolis, MN, 55455, USA}

\author{Doyal A. Harper}
\affil{Department of Astronomy and Astrophysics, University of Chicago, Chicago, IL 60637, USA}

\author{Thomas K. Henning}
\affil{Max Planck Institute for Astronomy, Koenigstuhl 17, D-69117 Heidelberg, Germany}

\author{Alex Lazarian}
\affil{Department of Astronomy, University of Wisconsin, Madison, WI 53706, USA}

\author{Leslie W. Looney}
\affil{Department of Astronomy, University of Illinois, 1002 West Green Street, Urbana, IL 61801, USA}

\author{Joseph M. Michail}
\affil{Center for Interdisciplinary Exploration and Research in Astrophysics (CIERA), and Department of Physics \& Astronomy, Northwestern University, 2145 Sheridan Rd, Evanston, IL, 60208, USA}
\affil{Department of Astrophysics and Planetary Science, Villanova University, 800 E. Lancaster Ave., Villanova, PA 19085, USA}
\affil{Department of Physics, Villanova University, 800 E. Lancaster Ave., Villanova, PA 19085, USA}

\author{Giles Novak}
\affil{Center for Interdisciplinary Exploration and Research in Astrophysics (CIERA), and Department of Physics \& Astronomy, Northwestern University, 2145 Sheridan Rd, Evanston, IL, 60208, USA}

\author{Fabio P. Santos}
\affil{Center for Interdisciplinary Exploration and Research in Astrophysics (CIERA), and Department of Physics \& Astronomy, Northwestern University, 2145 Sheridan Rd, Evanston, IL, 60208, USA}
\affil{Max-Planck-Institute for Astronomy, K\"onigstuhl 17, 69117 Heidelberg, Germany}

\author{Kartik Sheth}
\affil{NASA Headquarters, 300 E Street SW, DC  20546, USA}

\author{Javad Siah}
\affil{Department of Physics, Villanova University, 800 E. Lancaster Ave., Villanova, PA 19085, USA}

\author{Gordon J. Stacey}
\affil{Department of Astronomy, Cornell University, Ithaca, NY 14853, USA}

\author{Johannes Staguhn}
\affil{Dept. of Physics \& Astronomy, Johns Hopkins University, Baltimore, MD 21218, USA}
\affil{NASA Goddard Space Flight Center, Greenbelt, MD 20771, USA}

\author{Ian W. Stephens}
\affil{Center for Astrophysics $|$ Harvard \& Smithsonian, 60 Garden Street, Cambridge, MA, USA}

\author{Konstantinos Tassis}
\affil{Department of Physics and ITCP, University of Crete, GR-70013 Heraklion, Greece}
\affil{Institute of Astrophysics, Foundation for Research and Technology-Hellas, Vassilika Vouton, GR-70013 Heraklion, Greece}

\author{Christopher Q. Trinh}
\affil{USRA/SOFIA, NASA Armstrong Flight Research Center, Building 703, Palmdale, CA 93550, USA}

\author{John E. Vaillancourt}
\altaffiliation{Current address: Lincoln Laboratory, Massachusetts Institute of Technology, Lexington, Massachusetts 02421-6426}
\affil{Universities Space Research Association, NASA Ames Research Center, Moffett Field, CA 94035}
\affil{Enrico Fermi Institute, Department of Astronomy and Astrophysics, University of Chicago, Chicago, IL 60637, USA}

\author{Derek Ward-Thompson}
\affil{Jeremiah Horrocks Institute, University of Central Lancashire, Preston PR1 2HE, United Kingdom}

\author{Michael Werner}
\affil{NASA Jet Propulsion Laboratory, California Institute of Technology, 4800 Oak Grove Drive, Pasadena, CA 91109, USA}

\author{Edward J. Wollack}
\affil{NASA Goddard Space Flight Center, Greenbelt, MD 20771, USA}

\author{Ellen G. Zweibel}
\affil{Department of Astronomy, University of Wisconsin, Madison, WI 53706, USA}

\collaboration{(HAWC+ Science Team)}

\begin{abstract}

SOFIA HAWC+ polarimetry at $154~\micron$ is reported for the face-on galaxy M51 and the edge-on galaxy NGC 891. For M51, the polarization vectors generally follow the spiral pattern defined by the molecular gas distribution, the far-infrared (FIR) intensity contours, and other tracers of star formation. The fractional polarization is much lower in the FIR-bright central regions than in the outer regions, and we rule out loss of grain alignment and variations in magnetic field strength as causes. When compared with existing synchrotron observations, which sample different regions with different weighting, we find the net position angles are strongly correlated, the fractional polarizations are moderately correlated, but the polarized intensities are uncorrelated. We argue that the low fractional polarization in the central regions must be due to significant numbers of highly turbulent segments across the beam and along lines of sight in the beam in the central 3 kpc of M51. For NGC 891, the FIR polarization vectors within an intensity contour of 1500 $\rm{MJy~sr^{-1}}$ are oriented very close to the plane of the galaxy. The FIR polarimetry is probably sampling the magnetic field geometry in NGC 891 much deeper into the disk than is possible with NIR polarimetry and radio synchrotron measurements. In some locations in NGC 891 the FIR polarization is very low, suggesting we are preferentially viewing the magnetic field mostly along the line of sight, down the length of embedded spiral arms. There is tentative evidence for a vertical field in the polarized emission off the plane of the disk.

\end{abstract}


\section{Introduction} \label{sec:intro}

 A face-on and an edge-on galaxy each provides the observer with a unique advantage that enhances the study of the properties of spiral galaxies in general. For a face-on galaxy, there is far less confusion caused by multiple sources along the line of sight, a minimum column density of gas, dust and cosmic ray electrons, and a clear view of spiral structure. For an edge-on galaxy, the vertical structure of the disk is easily discernible, vertical outflows and super-bubbles can be seen, and the fainter, more diffuse halo is now more accessible. M51 and NGC 891 provide two well studied examples of nearly face-on (M51) and edge-on (NGC 891) galaxies. We are interested in probing the magnetic field geometry in these two systems to compare far-infrared (FIR) observations with optical, near-infrared (NIR) and radio observations, and to search for clues to the mechanism(s) for generating and sustaining magnetic fields in spiral galaxies.

Over the past few decades, astronomers have detected magnetic fields in galaxies at many spatial scales. These studies have been performed using optical, NIR, CO and radio observations \citep[see][for example]{kron94, zwei97, beck04, beck15, mont14, jone00, li11}. In most nearly face-on spirals, synchrotron observations reveal a spiral pattern to the magnetic field, even in the absence of a clear spiral pattern in the surface brightness \citep{flet10, beck04}. If magnetic fields are strongly tied to the orbital motion of the gas and stars, differential rotation would quickly wind them up and produce very small pitch angles. The fact that this is clearly not the case is an argument in favor of a decoupling of the magnetic field geometry from the gas flow due to diffusion of the field \citep{beck13}, which is expected in highly conductive ISM environments \citep[e.g.][]{laza12}.

Radio observations measure the polarization of centimeter (cm) wave synchrotron radiation from relativistic electrons, which is sensitive to the cosmic ray electron density and magnetic field strength \citep{jone74, beck15}. \cite{li11} measured the magnetic  field geometry in several star forming regions in M33 by observing  CO emission lines polarized due to the Goldreich-Kylafis effect \citep{gold81}, although there is an inherent $90\degr$ ambiguity in the position angle with this technique. Studies of interstellar polarization using the transmission of starlight at optical and NIR wavelengths can reveal the magnetic field geometry as a result of dichroic extinction by dust grains aligned with respect to the magnetic field \citep[e.g.,][]{jowh15} where the asymmetric dust grains are probably aligned by radiative alignment torques \citep{laza07, ande15}. However, polarimetric studies at these short wavelengths of diffuse sources such as galaxies can be affected by contamination from highly polarized, scattered starlight. This light originates with stars in the disk and the bulge, that subsequently scatters off dust grains in the interstellar medium \citep{jone12}. The optical polarimetry vector map of M51 \citep{scar87} was claimed to trace the interstellar polarization in extinction and does indeed follow the spiral pattern. As we will see later in the paper, it also demonstrates a remarkable degree of agreement with our HAWC+ map of the magnetic field geometry. A more recent upper-limit to the polarization measured at NIR wavelengths appeared to rule out dichroic extinction of starlight as the main polarization mechanism \citep{pave12}. The scattering cross section of normal interstellar dust declines much faster $(\sim \lambda^{-4}$ between 0.55 and $1.65~\micron$) than its absorption, which goes as $\sim \lambda^{-1}$ \citep{jowh15}. It is therefore possible that the optical polarization measured by \cite{scar87} is due to scattering, rather than extinction by dust grains aligned with the interstellar magnetic field, since polarimetric studies at these short wavelengths of diffuse sources such as galaxies can be affected by contamination from highly polarized scattered light \citep{wood97, seon18}. Nevertheless, the similarity we will find between the optical data and FIR results is striking, but if they are both indicating the same magnetic field, then the non-detection in the NIR is a mystery. Note that we will find a similar dilemma in comparing the optical and FIR polarimetry of NGC 891.

Observing polarization at FIR wavelengths has some advantages over, and is very complementary to, observations at optical, NIR and radio cm wavelengths for the following reasons. 1) The dust is being detected in polarized thermal emission from elongated grains oriented by the local magnetic field \citep[see the review by][]{jowh15}, not extinction of a background source, as is the case at optical and NIR wavelengths. 2) Scattering is not a contaminant since the wavelength is much larger than the grains, and much higher column densities along the line of sight can be probed. 3) Faraday rotation, which is proportional to $\lambda^2$, must be removed from radio synchrotron observations, and can vary across the beam, is insignificant for our FIR polarimetry \citep{krau66}. 4) The inferred magnetic field geometry probed by FIR polarimetry is weighted by dust column depth and dust grain temperature, not cosmic ray density and magnetic field strength, as is the case for synchrotron emission. In this paper we report observations at $154~\micron$ of both M51 and NGC 891 using HAWC+ on SOFIA \citep{harp18} with a FWHM beam size of 560 and 550 pc respectively. In all cases, we have rotated the FIR polarization vectors by $90\degr$ to indicate the implied magnetic field direction. This rotation is also made for synchrotron emission at radio wavelengths, but is $not$ made for optical and NIR polarimetry where the polarization is caused by extinction (unless contaminated by scattering), not emission, and directly delineates the magnetic field direction. The polarization position angles are not true vectors indicating a single direction, but the term `vector' has such a long historical use that we will use that term here to describe the position angle and magnitude of a fractional polarization at a location on the sky. The polarization is a true vector in a Q,U or Q/I,U/I diagram, but this translates to a $180\degr$ duplication on the sky.

\section{Far-infrared Polarimetric Observations} \label{sec:OBS}


The $154~\micron$ HAWC+ observations presented in this paper were acquired as part of SOFIA Guaranteed Time Observation program 70\_0609 and Director's Discretionary Time program 76\_0003.
The HAWC+ imaging and polarimetry -- resulting in maps of continuum Stokes I, Q, U -- used the standard Nod Match Chop (NMC) observing mode, performed at 4 half-wave plate angles and sets of 4 dither positions.  Multiple dither size scales were used in order to even the coverage in the center of the maps.



The M51 data were acquired during two flight series, on SOFIA flights 450, 452, and 454 in November 2017 and on flights 545 and 547 in February 2019.
The chop throw for the Nov. 2017 observations was 6.7 arcminutes at a position angle of 105 degrees east of north.  For the Feb. 2019 observations, the chop throw was 7.5 arcminutes in the east-west direction.
The total elapsed time for the M51 observations was 4.6 hours.  The observations with telescope elevation $> 58^\circ$ at the end of flight 547 were discarded due to vignetting by the observatory door.  Otherwise, conditions were nominal.


The NGC 891 data were acquired on flight 450 and on flights 506 and 510 in September 2018.  The chop throw for all observations was 5.0 arcminutes at a position angle of 115 degrees east of north.
The total elapsed time for the NGC 891 observations was 3.2 hours.  Four dither positions with telescope tracking problems during flight 450, which did not successfully run through the data analysis pipeline, were discarded.  Otherwise, observing conditions were nominal.


\subsection{Data Reduction}

All HAWC+ imaging and polarimetry were reduced with HAWC+ data reduction pipeline 1.3.0beta3 (April 2018).
Following standard pipeline practice, we subtracted an instrumental polarization $\{q_i, u_i\}$, calibrated with separate `skydip' observations, having a median value of $\sqrt{q_i^2 + u_i^2}$ of 2.0\% over the detector array.
The final uncertainties were increased uniformly by $\sim30-40\%$ based on the $\chi^2$ consistency check described by \citet{Santos19}.  We applied map-based deglitching as described by \citet{chus19}. Due to smoothing with a kernel approximately half the linear size of the beam, the angular resolution in the maps (based on Gaussian fits) is 14\arcsec\ FWHM at 154 \micron. Since both galaxies are well out of the Galactic plane, reference beam contamination is minimized.

The flux densities in the maps were calibrated using observations of Solar System objects, also in NMC mode.  Due to the lack of a reliable, calibrated SOFIA facility water vapor monitor at the time of the observations, the version 1.3.0 pipeline uses an estimate of far-IR atmospheric absorption that is dependent on observatory altitude and telescope elevation, but is constant in time.  For all observations, we used the default pipeline flux calibration factor, for which we estimate 20\% absolute uncertainty.  For each galaxy, the maps from the two flight series, analyzed separately, show flux calibration consistency to within 5\% .  For M51, we adjusted the coordinates of the Feb. 2019 map (with a simple translation in both axes) prior to coaddition with the Nov. 2017 map.  The relative alignment of the per-flight-series maps for NGC 891 was within a fraction of a beam without adjustment.

Alignment of the coordinate system for M51 supplied by the pipeline was checked against VLA 3.6cm, 6.2cm, and 20.5cm \citep{flet11}, Spitzer $8~\micron $\citep{smith07}, and Herschel $160~\micron$ maps \citep{pilb10}. We did this by matching 6 small, high surface brightness regions between our $154~\micron$ map and the maps at the other wavelengths. We found that the HAWC+ map was consistently $4\pm 1\arcsec$ south relative to the comparison maps. For this reason, we have added an offset of $4\arcsec$ N to our maps of M51. Since we are not making any comparisons of NGC 891 with high resolution maps at other wavelengths, we made no adjustment to the coordinate system for that galaxy.



\subsection{Polarimetry Analysis} \label{sec:Analysis}

For both galaxies we computed the net polarization in different synthetic aperture sizes, depending on the signal-to-noise (S/N) in the data. The pixel size is $3.4 \arcsec$, or $\sim$ 1/4 a FWHM beam width. In all cases we used the I, Q and U intensity and error maps to form the polarization vectors. The results reported here were obtained by placing different sized synthetic apertures on the images and computing intensities from the sums of individual pixels and the errors from the sums of the error images in that aperture in quadrature. The errors and intensities in the individual pixels are not statistically independent, since they were created by combining intermediate images in the data processing and then smoothed with a truncated Gaussian with FWHM = 2.04 pixels $(6.93\arcsec)$. We determined the effect of the Gaussian kernel on the computed errors by applying it to maps with random noise. As a result of this exercise, we increased the computed error by factors of 1.69 for the $2\times2$ pixel (half beam), 2.27 for the $4\times4$ pixel (one beam) and 2.56 for the $8\times8$ pixel (two beam) synthetic apertures. 

An additional concern is spatially correlated noise such as might be due to incomplete subtraction of atmospheric noise and other effects.  A thorough investigation into the possibility of correlated noise in our data is beyond the scope of this paper and will be addressed in a later paper, but we report the results of a simple test for spatially correlated noise carried out by the HAWC+ instrument team (Fabio P. Santos) in 2017 on B and C observations of HL Tau. This analysis showed that an approximate quadrupling of the sky area being combined causes the noise in the data (compared to what would be expected from uncorrelated noise) to increased by a factor of 1.06.  Specifically, results were compared for a Gaussian smoothing kernel of $4\arcsec$ FHWM truncated at an $8\arcsec$ diameter and one having $7.8\arcsec$ FHWM with truncation at a $15.6\arcsec$ diameter. 

For this reason we have made extra cuts in Stokes I (total intensity) at a S/N of 50:1 for M51 and 30:1 for NGC 891, and increased the error for the largest synthetic aperture of $8\times8$ pixels by a factor of 1.06. We are particularly concerned about the scientifically important inter-arm and halo regions, which have low intensity and need to use the larger synthetic aperture. Q and U are $intensities$, and small spurious values will adversely influence the net polarization derived for regions of low intensity, but not high intensity. For example, at a contour level of 100 $\rm{MJy~sr^{-1}}$ between the arms, a 1 $\rm{MJy~sr^{-1}}$ value for Q that is due to a glitch, a bad pixel, or residual flux from image subtraction will produce a 1\% polarization that is not real. In the arm where the intensity is $\sim 800$ $\rm{MJy~sr^{-1}}$, this would contribute no more than 0.12\%. The final computed polarization was then corrected for polarization bias \citep{Wardle74, Sparks99} and cuts were made at a fractional polarization for a final S/N of $ \geq 3:1 $ and S/N between 2.5:1 and 3:1. 

To further guard against systematic errors in the I, Q and U maps at lower intensities, we made a cut using the total intensity error I$_{err}$ map at $\sigma > 0.003$ Jy/pixel. This removed the outer regions of the images where there was incomplete overlap in the dithered images. This final cut made little difference in the M51 polarimetry results where less than 10\% of the image was removed. But, for NGC 891 about 20\% of the image was removed and the northern and southern extremes of the disk in NGC 891 were excluded. Note that the edge-on disk in NGC 891 is at least $10\arcmin$ long, and our HAWC+ image spans only about $5\arcmin$ along the disk, centered on the nucleus. In an upcoming paper we will be working with existing and new HAWC+ data on M51 and will create smoothed images starting with the raw data.

\section{M51}\label{sec:M51}

\subsection{Introduction}

M51 is not only a face-on spiral galaxy but also a two-arm, grand design spiral \citep[e.g.][]{rand92}, at a distance of 8.5 Mpc \citep{mcqu16}. It is clearly interacting with M51b and tails and bridges in the outer regions of the two galaxies are shared, while in the inner regions of M51 the spiral structure appears to be unaffected by the companion. Our observations did not reach far enough from the center of the galaxy to include M51b. Because of its low inclination, M51 shows well defined spiral arms and well separated arm and inter-arm regions. This makes M51 an excellent laboratory to study how the magnetic field geometry changes from arm to inter-arm regions due to the effect of spiral density waves and turbulence.  Star formation in M51 is located mostly in the spiral arms and in the central region, but some gas and star formation are also detected in the inter-arm regions \citep[e.g.,][]{koda09}. Molecular gas is strongly correlated with the optical and infrared spiral arms and shows evidence for spurs in the gas distribution \citep{schi17}. The magnetic field geometry M51 was studied at radio wavelengths by \cite{flet11}, who find that the overall geometry revealed in the polarization vectors follows the spiral pattern, but there is depolarization in their larger $15\arcsec$ 20.5 cm beam. They find that the 6.2 cm polarized emission is probably strongly affected by sub-beam scale anisotropies in the field geometry. Our HAWC+ observations allow us to study the magnetic field geometry as measured by dust emission instead of cosmic ray electrons, and thereby sample the line of sight differently, and also probe denser components of the ISM than is possible at optical and NIR wavelengths. 

\subsection{Magnetic Field Geometry}

\begin{figure}
    \includegraphics[width=\columnwidth]{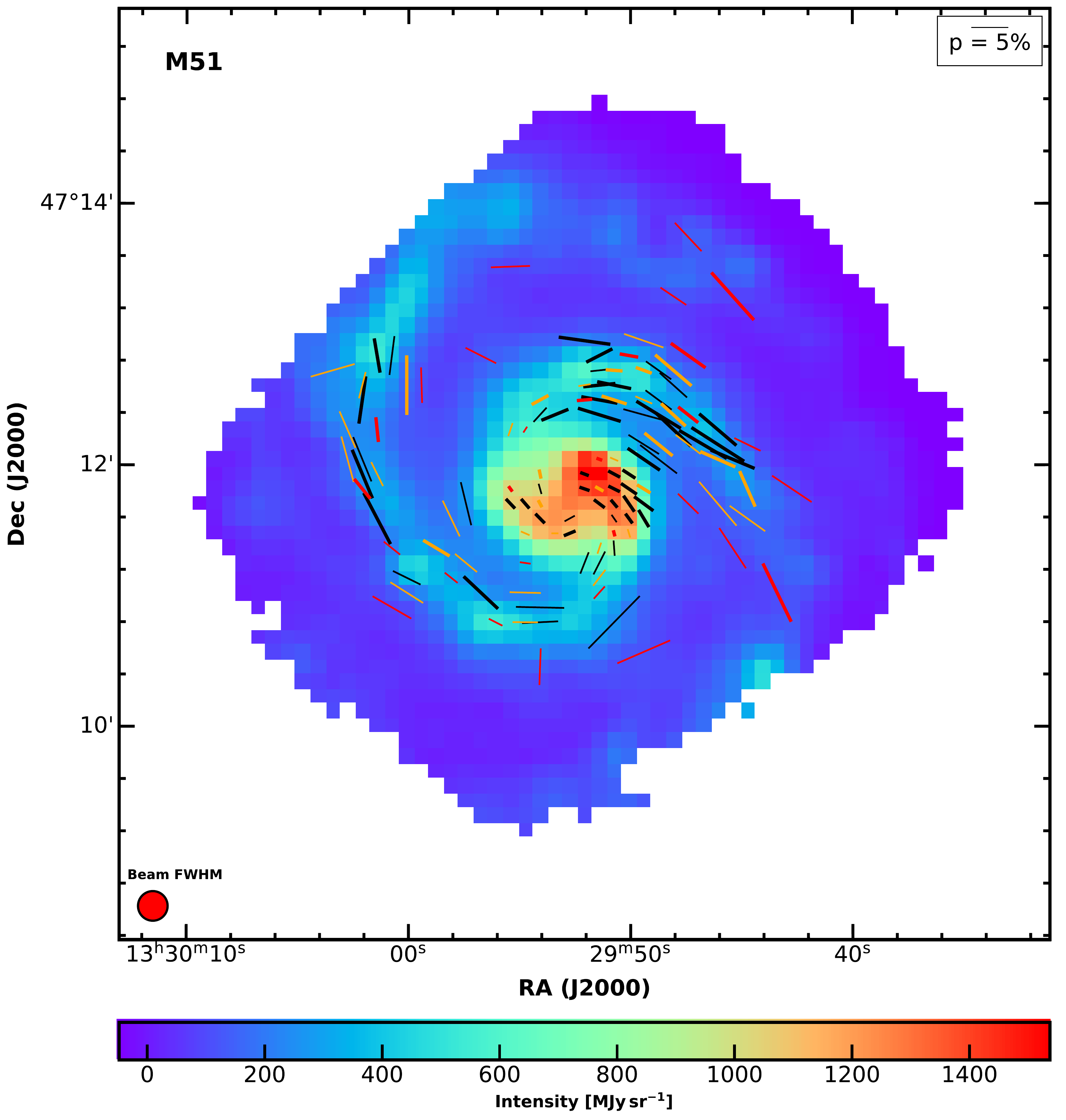}
    \caption{\label{fig:Map_final} Fractional polarization vector map of M51 at a wavelength of $154~\micron$, with the vectors rotated $90\degr$ to represent the inferred magnetic field direction. Data points using a square $6.8 \arcsec \times 6.8 \arcsec$ `half' beam are plotted in black. Data points using a $13.6 \arcsec \times 13.6 \arcsec$ `full' beam are plotted in orange, and red vectors are computed using a $27.2 \arcsec \times 27.2\arcsec$ square beam. The red disk in the lower left corner indicates the FWHM footprint of the HAWC+ beam on the sky at $154~\micron$. Colors in the underlying image define the $154~\micron$ continuum intensity. Vectors with S/N $\geq 3:1$ have thick lines and vectors with S/N from 2.5:1 to 3:1 have thin lines.}
\end{figure}

\begin{figure}
    \includegraphics[width=\columnwidth]{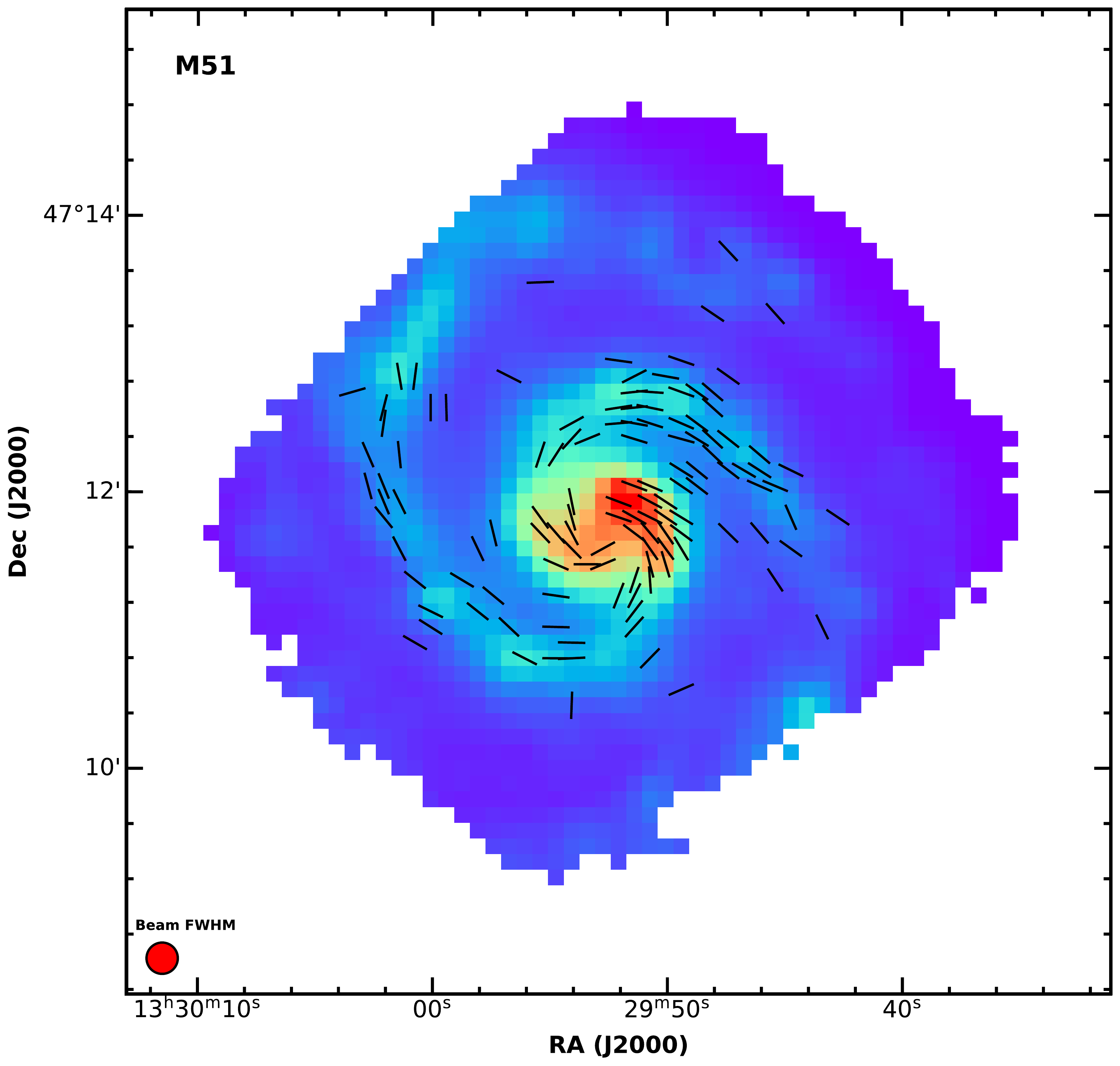}
    \caption{\label{fig:Map_final_constlength} Same as Figure \ref{fig:Map_final}, except all of the polarization vectors have been set to the same length and color to better illustrate their position angles.}
\end{figure}

The polarization vector map of M51 is shown in Figure \ref{fig:Map_final}, where the polarization vectors have been rotated $90\degr$ to show the inferred magnetic field geometry. Fractional polarization values range from a high of 9\% to a low of 0.6\%, about $3\sigma$ above our estimated limiting fractional polarization of 0.2\% \citep{jone19}. Clearly evident in Figure 1 is a strong correlation between the position angles of the FIR polarimetry and the underlying spiral arm pattern seen in the color map. This can be better visualized in Figure \ref{fig:Map_final_constlength}, where all the polarization vector lengths have been set to unity, and only the position angle (PA) is quantified. 

In spiral galaxies, the spiral pattern is often fitted with a logarithmic spiral \citep[e.g.][]{Seigar98, davi12, a mathematical curve that is characterized by a constant pitch angle. The pitch angle is an empirical parameter that quantifies the morphology of galaxies regardless of their distance. Pitch angles for the spiral features in M51 have been investigated at different wavelengths and using different methods.} \cite{Shetty07} found a pitch angle of 21.1$^{\circ}$ for the bright CO emission in the spiral arms. \cite{Hu13} suggested 17.1$^{\circ}$ and 17.5$^{\circ}$ for each of the two arms using SDSS images, and \cite{Puerari14} determined the pitch angle of 19$^{\circ}$ for the arms from $8~\micron$ images. Also, several investigators find that the pitch angles are variable depending on the location \citep[e.g.,][]{HowardByrd90, Patrikeev06, Puerari14}.

\begin{figure}
    \centering
    \includegraphics[width=.6\columnwidth]{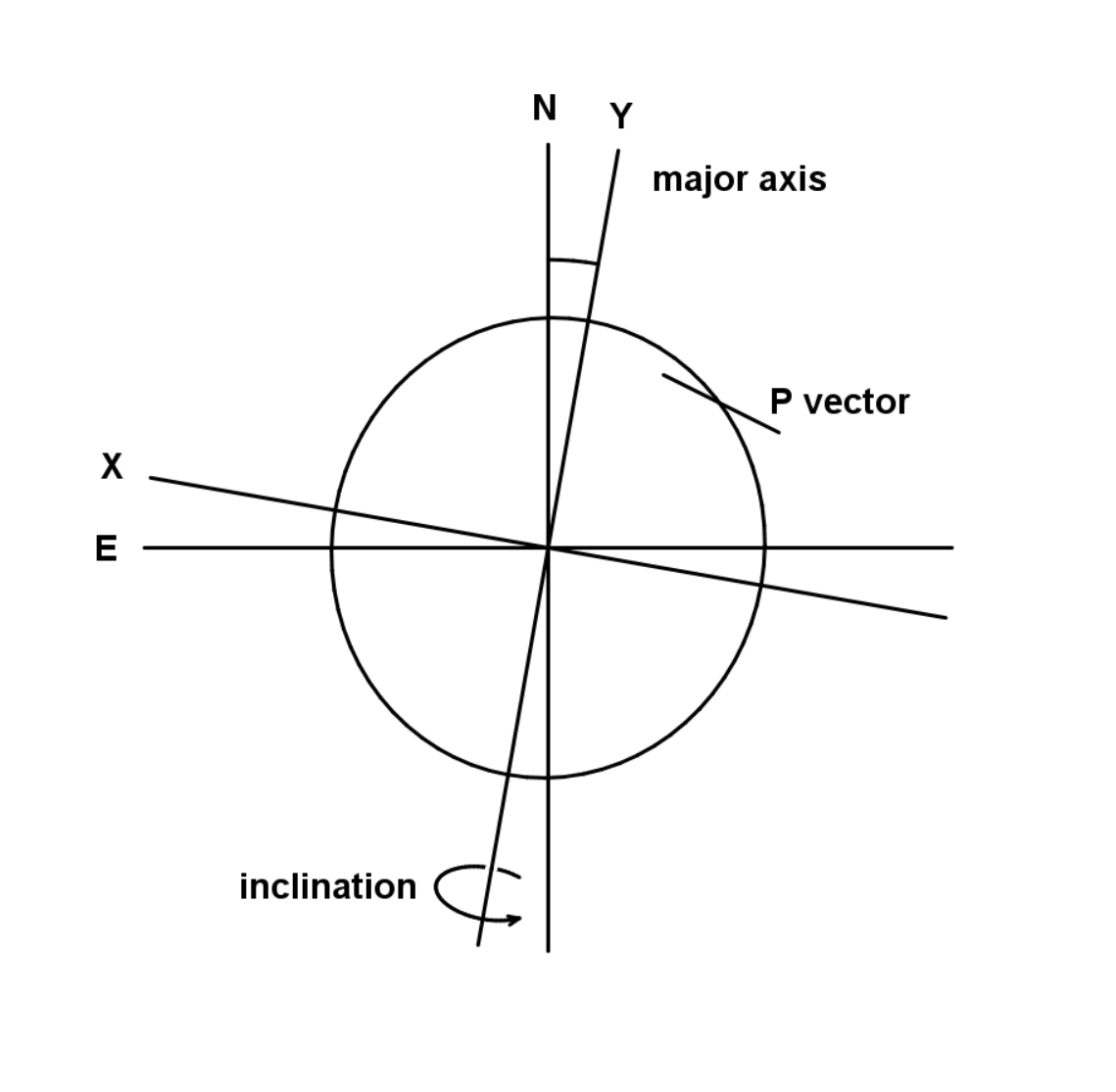}
    \caption{\label{fig:geo} Geometry used to de-project the polarization vectors so that their individual pitch angles can be calculated. The inclination with respect to the plane of the sky is $20\degr$ and the major axis (labeled Y) of the ellipse (a circle in projection) is  $170\degr$ east of north. We are assuming the magnetic field vectors in the disk of M51 have $\textit{no}$ vertical component when computing the de-projection. The polarization vector is shown relative to a circle (in projection), which has a pitch angle of zero.}
\end{figure}

M51 is not perfectly face-on, but rather is tilted to the line of sight. \cite{Shetty07} used the values for the inclination of $20\degr$ and a position angle for the major axis of $170\degr$ from \cite{tull74} in their analysis of the spiral arms seen in CO emission. This geometry is illustrated in Figure \ref{fig:geo}. Using these same parameters and assuming the intrinsic magnetic field vector has $\textit{no}$ component perpendicular to the disk, we can de-project our vectors and compute their individual pitch angles using the geometry from Figure \ref{fig:geo} \citep[see][]{lope19}. Having de-projected our vectors, we can compare the pitch angles of our vectors with the pitch angle(s) of a model spiral where we compute $\Delta \theta  = \rm{PA}_{\rm{FIR}} - \rm{PA}_{\rm{spiral}}$ where PA indicates pitch angle for the (de-projected) FIR polarimetry vectors and the model spiral respectively. 

\begin{figure}
    \centering
    \includegraphics[width=.5\columnwidth]{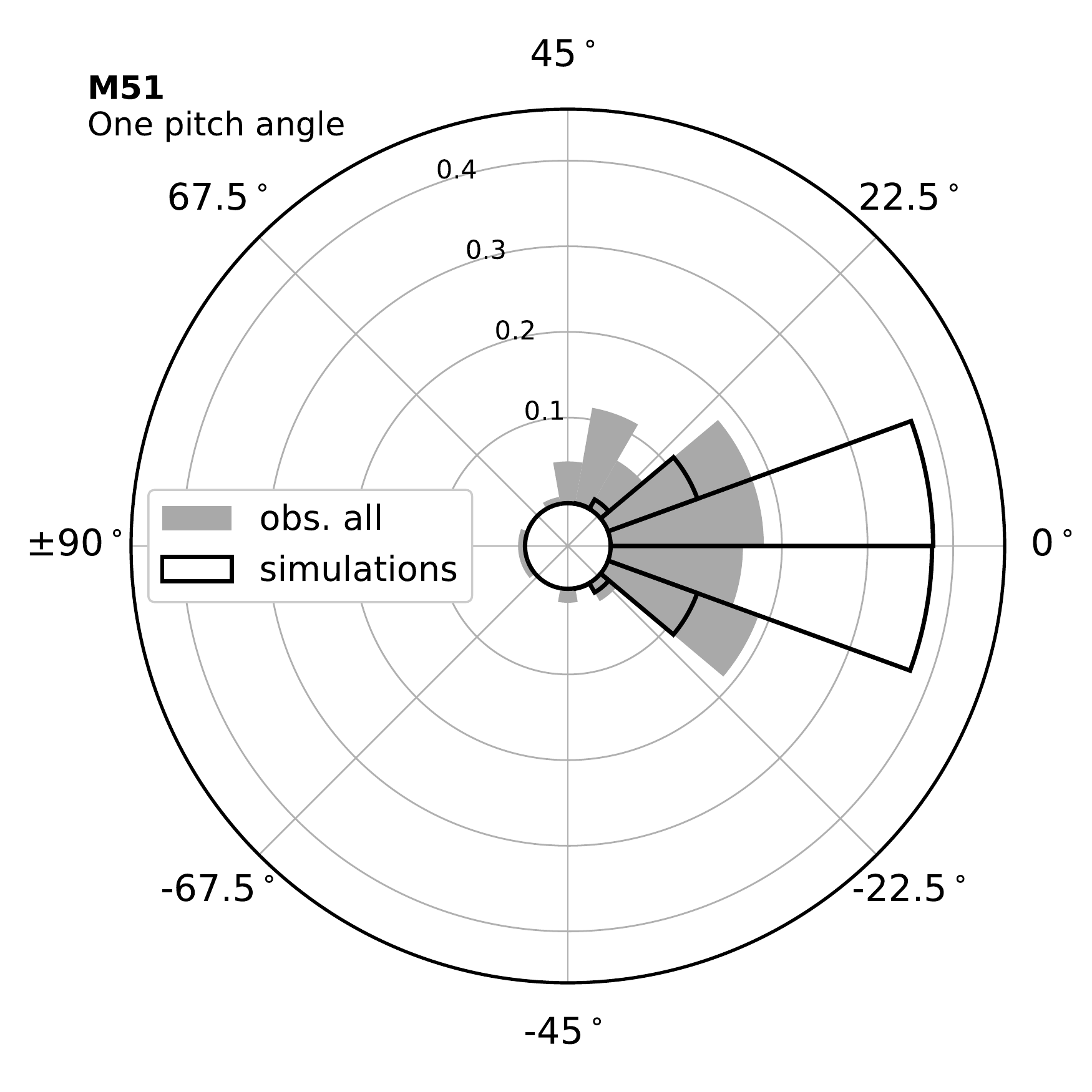}
    \caption{\label{fig:hist_onepitch_dev} Histogram distribution of $\Delta \theta$ between the pitch angle of our polarization vectors and a single pitch angle for the spiral arms of $21.1\degr$ (Model 1). Radial distance is the fraction of the total number of measurements. The area in grey shows the actual data and the solid lines show a simulation (see text) under the assumption that the pitch angles are intrinsically the same, and only errors in the data contribute to the dispersion.}
\end{figure}

First, we assume a single pitch angle of 21.1$^{\circ}$ from the CO observations for the model spiral arms, and compute $\Delta \theta$. We will call this Model 1. A normalized histogram of $\Delta \theta$ is shown in Figure \ref{fig:hist_onepitch_dev}. We simulated the expected distribution in $\Delta \theta$ under the assumption that the vectors and the spiral arm pitch angle were the same, and only errors in the FIR polarization data were responsible for the dispersion in the angle difference. We generated simulated data assuming the errors in polarization position angle are Gaussian distributed for each vector and ran a Monte Carlo routine that generated simulated distributions, repeating 1000 times. Since the simulated data are assumed to follow the arm exactly, the peak of the distribution function is set at $\Delta \theta = 0$. When the observational data and simulation are compared, the distribution of observed $\Delta \theta$ is broader than the simulated one with a standard deviation of $\sigma = 23\degr$ compared to $\sigma = 9\degr$ for the simulation. The observational data shows greater departure from a single pitch angle than can be accounted for by errors in the FIR polarimetry vector position angles alone.

\begin{figure}
    \includegraphics[width=\columnwidth]{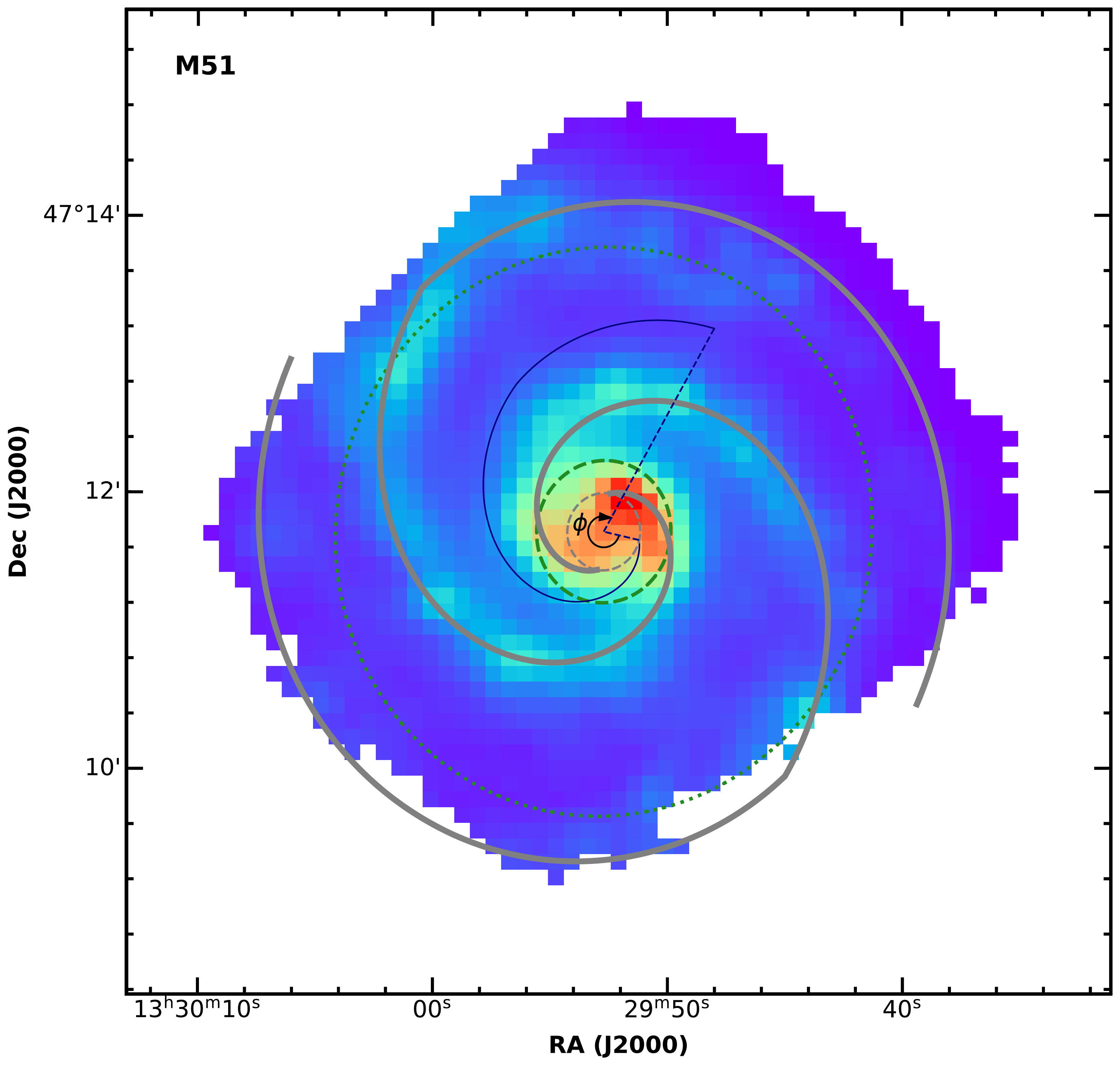}
    \caption{\label{fig:M51_spiral} Model 2 geometry using two spiral arm pitch angles (shown in grey) that we used to compute the distribution of $\Delta \theta$ for this case. The inner part has the pitch angle of $21.1\degr$, and the outer part a pitch angle of 3.9$^{\circ}$. The green dashed and dotted lines are the inner resonance and the co-rotation radii respectively, described in \cite{tull74}. The angle $\phi$ is used to define a measure of distance along a spiral `feature'. That is, we assume the basic two pitch angle model (shown in grey) extends between the arms (shown in blue).}
\end{figure}

\begin{figure}
    \includegraphics[width=\columnwidth]{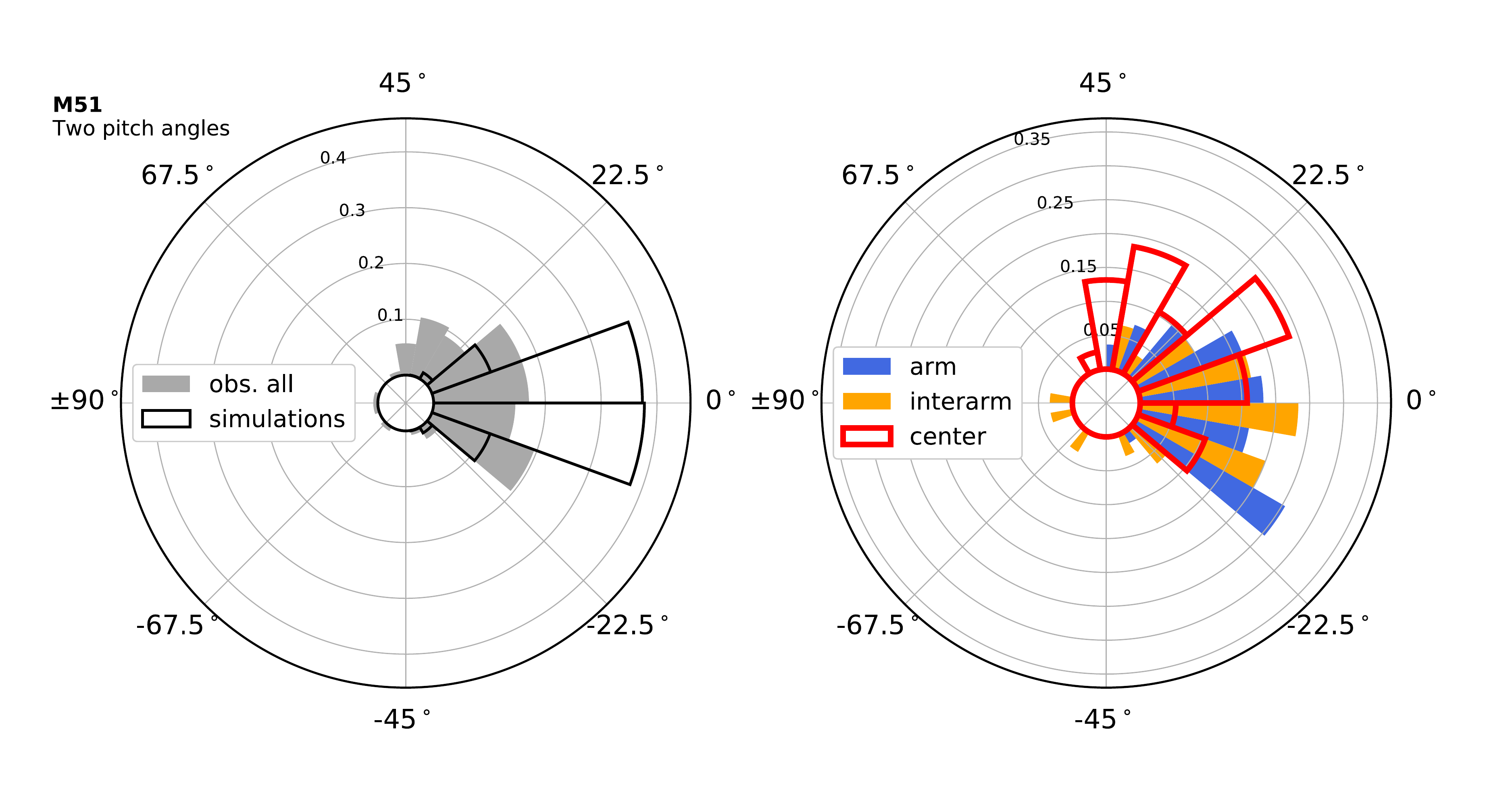}
    \caption{\label{fig:hist_twopitches_dev} Distribution of $\Delta \theta$ as in Figure \ref{fig:hist_onepitch_dev}, but using Model 2, which has two pitch angles. Grey and black represent the simulation and observations respectively. In the right hand figure, the observation are subdivided into arm, inter-arm, and central regions (see text), which are indicated by blue, orange, and red color, respectively. The locations of the different regions are defined in \cite{pine18}. Although very similar in appearance, the left panel is not identical to Figure \ref{fig:hist_onepitch_dev}}
\end{figure}

Next we modeled the spiral features with two pitch angles, with a change in pitch angle chosen to fit the FIR intensity data by eye. We will call this Model 2. The resulting model spiral arms are shown in Figure \ref{fig:M51_spiral} where the inner spiral arms at a radial distance of $137\arcsec$ from the center retain the 21.1$^{\circ}$ pitch angle based on the CO observations for part of the arms, and then a much tighter pitch angle of 3.9$^{\circ}$ is used for the outer arms. Following the same procedure as before, we computed the angle difference between the pitch angles of the polarization vectors and the spiral arms and ran a simulation of these differences, assuming they are intrinsically the same, and only observational errors are responsible for the dispersion in the differences. For this two pitch angle case, the results are plotted in Figure \ref{fig:hist_twopitches_dev}. Even with the two pitch angle model, the dispersion in $\Delta\theta$ is much greater than can be accounted for by the observational errors with nearly identical standard deviations to Model 1. To explore the spiral pattern in our polarimetry vectors in more detail, we separated the magnetic field vectors into arm, inter-arm, and center regions. These regions are classified according to the mask given in Figure 1 of \cite{pine18}, where the center region is roughly the inner 3 kpc (in diameter). Note that we are interpolating both models into the inter-arm region (see the blue line in Figure \ref{fig:M51_spiral}). The distribution of $\Delta\theta$ for these separate regions is shown in the right hand panel of Figure \ref{fig:hist_twopitches_dev}. The vectors in the center group have a distinct positive mean offset of $17.4\degr$, which means a more open spiral pattern compared to the model pitch angle. The inter-arm and arm groups have no clear offset from zero, but the dispersion is still much larger than can be explained by measurement errors alone.

\begin{figure}
\centering
    \includegraphics[width = \columnwidth]{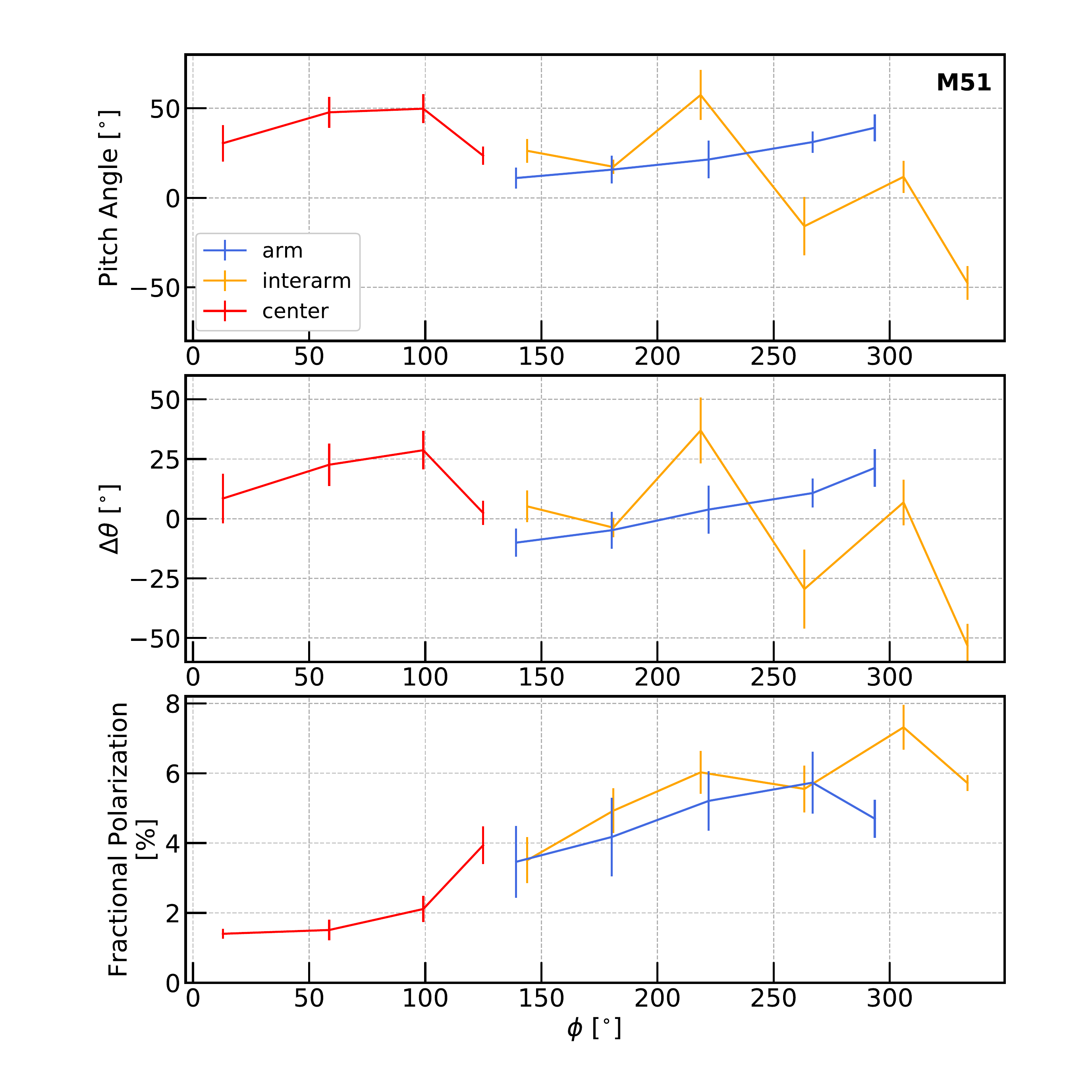}
    \caption{\label{fig:PitchPvsphase} Pitch angle of the FIR vectors (top), the deviation of these pitch angles from the spiral arms (middle), and the fractional polarization (bottom) depending on $\phi$, an angular distance along the arm defined in \ref{fig:M51_spiral}, assuming Model 2 with the two pitch angles for the spiral arms. Vertical bars represent the standard deviation of the data within each bin, not an error in measurement. Red, blue, and orange represent the center, arm, and inter-arm group, respectively.}
\end{figure}

 In Figure \ref{fig:M51_spiral} we define $\phi$, a measure of the angular distance along a spiral feature, increasing from zero clockwise around the galaxy (along the spiral features). We define a spiral feature for each point in the map  (see Figure \ref{fig:M51_spiral}), and extrapolate back to the central region to determine the angular distance $\phi$.  The pitch angle, averaged over intervals of $\phi=40\degr$, as a function of angular distance along a spiral model line, is illustrated in Figure \ref{fig:PitchPvsphase}. The top panel is the pitch angle of the FIR polarization vectors. The middle panel plots $\Delta \theta$, the difference between Model 2 and observed pitch angles. The lower panel shows the trend in fractional polarization with $\phi$. We find no statistically significant difference in the trends of fractional polarization with $\phi$ when comparing the arm and interarm regions. The dispersion for $\Delta\theta$ in the inter-arm region is large, and departs from the trend seen in the arm in the last data bin.
 
 Overall our FIR vectors follow the spiral arms in M51, but with fluctuations about the spiral arm direction that are greater than can be explained by measurement errors alone. \cite{step11} found no correlation between the magnetic field geometry in dense molecular clouds in the Milky Way and Galactic coordinates, and this may add a random component to the net position angles we are measuring in our large 560 pc beam. However, the relative contributions of emission from dense $(n_H > 100~\rm{cm}^{-3})$ and more diffuse regions in M51 to our $154\micron$ flux has not been modeled. The FIR vectors in the central region indicate a more open spiral pattern than seen in the molecular gas \citep{Shetty07}, opposite to what one would expect if the magnetic fields were wound up with rotation. Although our data in the inter-arm region are relatively sparse, the fractional polarization is statistically similar to the that in the arms, which are delineated by a higher FIR surface brightness.
 
 \cite{houd13} used the position angle structure function \citep{kobu94, hild09, houd16} to characterize the magnetic turbulence in M51 using the radio polarization data from \cite{flet11}. See section 3.4 for a comparison with the radio data. Analyzing the galaxy as a whole and using a 2D Gaussian characterization of the random component to the magnetic field, they found the turbulent correlation scale length parallel to the mean field was $98 \pm 5$ pc and perpendicular to the mean field was $53 \pm 3$ pc. This indicates that the random component has an anisotropy with respect to the spiral pattern, and could be interpreted as due to shocks in the spiral arms \citep{pine20} compressing anisotropic turbulence in a particular direction \citep{beck13}.  We will explore the position angle structure function in a later paper with new SOFIA/HAWC+ observations that will allow us to measure fainter regions due to increased integration time.
 
 \cite{houd13} also found that the ratio of random to ordered strengths of the magnetic field was tightly constrained to $\rm{B}_r/\rm{B}_o = 1.01 \pm 0.04$,  and this ratio is consistent with other work \citep[e.g.,][]{jkd, mivi08}. Assuming the spiral pattern represents the geometry of the ordered component,  the addition of a random component may explain our broad distribution of position angles with respect to the spiral structure.  Broadening of the distribution of $\Delta\theta$ by a random component depends on the number of turbulent segments in our beam. If we use the 100 pc turbulent correlation scale determined by \cite{houd13}, there are $> 25$ segments in our beam, which will largely 'average out' relative to the ordered component (see Figure 8 in \cite{jkd}). A simple broadening of the distribution due to this spatially small random component would not produce the number of position angles differing by  $60-90 \degr$ from the spiral pattern seen in Figure \ref{fig:hist_twopitches_dev}.  However, all of the vectors that depart by more than $60\degr$ are in the inter-arm region and  have S/N only between 2.5:1 and 3:1. The distribution of $\Delta\theta$ for the arm region (only) is much more similar to the simulation, with a mean value of only $5\degr$. The dispersion, however, is still a factor of 2 greater. Given the uncertainty in the contribution of a random component to the magnetic field, the FIR vectors in the arms (blue colored bars in Figure \ref{fig:hist_twopitches_dev}) could be consistent with the spiral pattern we defined in Figure \ref{fig:M51_spiral}, but without a better determination of the turbulent component, we can not make a better determination. Even with these uncertainties, there  remains a clear shift in the mean pitch angle for the center region to a more open (greater pitch angle) pattern than seen in the CO and star formation tracers. More sensitive observations, in particular for the inter-arm region, will be necessary to better define the correlation between the FIR vectors and the spiral pattern. 
 
Using broadband 20 cm observations with the VLA, \cite{mao15} studied the rotation measures in M51 in detail. They find that at 20 cm most of the observations are consistent with an external uniform screen (halo) in front of the synchrotron emitting disk. The disk itself produces synchrotron emission that is partially depolarized on scales smaller than 560 pc (which is our beam size), with most of the polarized flux originating in the top layer of the disk, then passing through the halo. The scale length for the rotation measure structure function in the halo is 1 kpc, which is consistent with blowouts and superbubbles from activity in the disk. Our FIR observations are tied to the warm dust in the disk and are largely insensitive to the magnetic field geometry in the halo, but should be sensitive to the formation of superbubbles which have their origin in the disk. We will be exploring the position angle pattern in more detail in a later paper.

\subsection{Polarization -- Intensity relation} \label{sec:PvsI} 

\begin{figure}
    \includegraphics[width=\columnwidth]{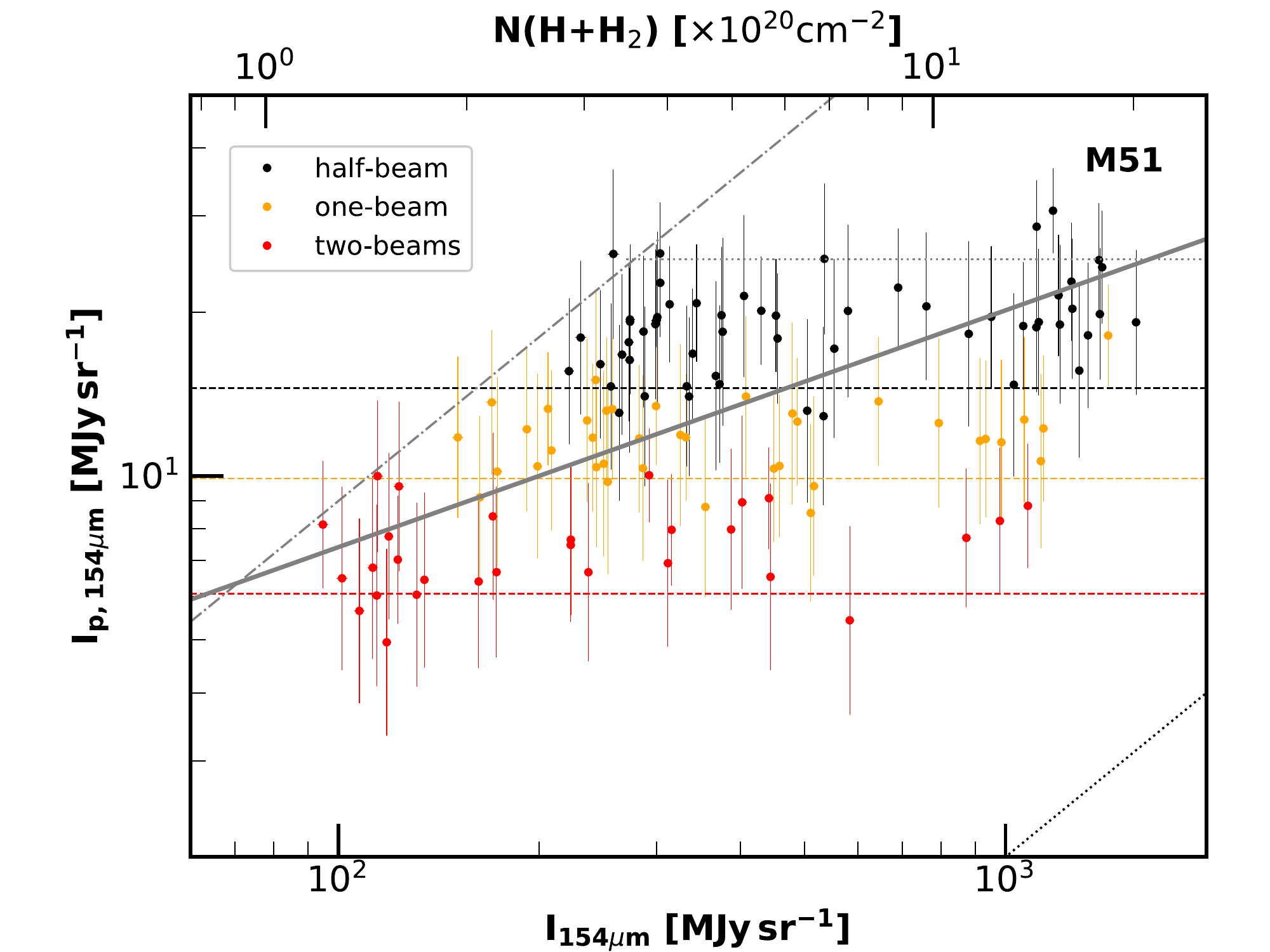}
    \caption{\label{fig:M51_IpI} The debiased polarized intensity plotted against the intensity at our wavelength of $154~\micron$ and derived hydrogen column depth (see text). The vector data shown in Figure \ref{fig:Map_final} were used. The grey solid line is a linear fit to the data with a slope of $\log \rm{I}_{\rm{p}154~\micron} = 0.43 \log \rm{I}_{154~\micron}$ ($\alpha=-.57$) calculated by an orthogonal distance regression (ODR) weighted by the squares of errors using \texttt{scipy.odr} module. Each dashed line of different color represents the $2.5\sigma$ observation limit estimated from the errors in Q and U in each bin size. The grey dash-dotted line in the upper left corner shows the maximum value of $\rm{I_p}$ corresponding to a maximum fractional polarization of $9\%$ (see text), and has a slope of +1.0 ($\alpha = 0$). The horizontal dotted line corresponds to an empirical upper boundary seen in the data at $\rm{I_p} = 25~ \rm{MJy~sr^{-1}}$ and corresponds to $\alpha = -1$. Finally, the line in the lower right hand corner shows the estimated $\pm 0.2\%$ limit in fractional polarization precision we can achieve with HAWC+ polarimetry \citep{jone19} in an ideal data set.} 
\end{figure}

In our previous FIR polarimetry of galaxies \citep{jone19, lope19} we found that the fractional polarization declines with intensity and column depth, and can often be characterized by a power law dependency $\rm{p} \propto \rm{I}^\alpha$. This trend is also common in the Milky Way \citep[e.g.,][]{plan15}, in particular in molecular clouds, and is commonly plotted as $\log(\rm{p})$ vs. $\log(\rm{I})$ \citep[e.g.,][]{fiss16, jone15, gala18, chus19}. In our previous papers we have used fractional polarization $\rm{p}$, but because of selection effects due to intensity cuts, the minimum measurable fractional polarization and a physical maximum in the fractional polarization are difficult to discern in that type of a plot. Instead, here we adopt plotting the polarized intensity $\rm{I_p}$ as a function of intensity or column depth. For comparison, a slope of $\alpha = -0.5$ in $\log(\rm{p})$ vs. $\log(\rm{I})$ (or column depth) is equivalent to a slope of $+0.5$ in $\log(\rm{I_p})$ vs. $\log(\rm{I})$. This can easily be seen through the relation $\rm{I_p}=\rm{pI}$.

For M51, this comparison is shown in Figure \ref{fig:M51_IpI}. The column density was computed assuming a constant temperature for the dust, and is therefore a simple multiplicative factor of the intensity. We used an emissivity modified blackbody function assuming a temperature of 25K \citep{Benford08}. The dispersion in derived temperature found using Herschel data was only $\pm 1.0$K, confirming that variation in temperature across M51 will not affect our results. We define an emissivity, $\epsilon$, which is proportional to $\nu^{\beta} $ using a dust emissivity index, $\beta$, of 1.5 from \cite{Boselli12}. We made use of the relation of the hydrogen column density, $\rm{N(H + H_2)} = \epsilon / (\textit{k} \mu \rm{m_H})$, with the dust mass absorption coefficient, $k$, of $0.1\ \rm{cm}^{2}~\rm{g}^{-1}$ at 250 $\micron$ \citep{Hildebrand83}, and the mean molecular weight per hydrogen atom, $\mu$ of 2.8 \citep{Sadavoy13}. The maximum expected fractional polarization of $9\%$ at $\sim 150~\micron$ is taken from \cite{hild95} and is within the range of dust models computed by \cite{guil18} that were based on Planck observations. This upper limit nicely delineates the boundary seen in the maximum $\rm{I_p}$ measured at low column depths in M51. 

Note that the lowest polarized intensities are associated with the larger $27.2\arcsec \times 27.2\arcsec$ aperture (labeled two-beam), and averaging over this aperture could artificially reduce the computed polarization if there is significant variation in position angle of the ordered component (not the random component) to the field within the aperture. However, even a $45\degr$ variation in position angle for the ordered component across the aperture would only reduce the net polarization by $1/\sqrt{2}$, yet the mean for the two-beam $\rm{I}_p$ is at least a factor of 3 lower than for the half-beam data. Also, the large aperture results are concentrated well away from the nucleus where the spatial variation in position angle is less. The primary cause of the vertical separation between the different beam sizes in Figure \ref{fig:M51_IpI} is S/N, rather than beam averaging. A simple linear fit (in log space) to all of the data in Figure \ref{fig:M51_IpI} has a slope less than $+0.5$. This translates to a slope more negative than $\alpha = -0.5$ in a $\log(\rm{p})$ vs. $\log(\rm{I})$ plot. Note that selection effects such as our minimum detectable polarized intensity are easy to delineate in Figure \ref{fig:M51_IpI}, as shown by the horizontal lines. Due to concerns about the effect the minimum detectable fractional polarization on the data points in the lower right of Figure \ref{fig:M51_IpI}, we will concentrate on examining the upper envelope of the data rather than the best-fit slope.

The upper limit in Figure \ref{fig:M51_IpI} has a slope of +1 (p = constant) up until $\rm{N(H + H_2)} \sim 3.5\times 10^{20}~\rm{cm^{-2}}$. The slope then changes and becomes flat ($\rm{I_p}$ = constant), and $\rm{I_p}=25~\rm{MJy~sr^{-1}}$ at greater column depth. This flat slope corresponds to a slope of $\alpha=-1$, as discussed above. For M51, the change in slope for the upper limit in polarized intensity occurs at approximately 1/3 the value of $\rm{N(H + H_2)} \sim  10^{21}~\rm{cm^{-2}}$ found by Planck for polarization in the Milky Way (see Figure 19 in \cite{plan15}). As mentioned above, a strong decline in fractional polarization with column density was also found for FIR polarimetry of M82, NGC 253 \citep{jone19} and NGC 1068 \citep{lope19}. Note that NGC 1068 has a powerful AGN which could create a more complex magnetic field, but most of the FIR polarimetry samples only the much larger, surrounding disk. \cite{lope19} suggested three possible explanations for the decline in fractional polarization with column depth, assuming the emission is optically thin. Polarization may be reduced if there are segments along the line of sight where 1) the grains are not aligned with the magnetic field, 2) the polarization is canceled because of crossed or other variations of the magnetic field on large scales, or 3) there are sections along the line of sight that contain turbulence on much smaller scale lengths than in lower column density lines of sight, contributing total intensity, but little polarized intensity. \cite{lope19} considered the contribution of regions that are sufficiently dense that their higher extinction may prevent the radiation necessary for grain alignment from penetrating. These regions make a very small a contribution to the FIR flux in the HAWC+ beam, simply because they are small in angular size and very cold. Although these dense cores probably experience a loss of grain alignment, they cannot have any effect on our observations of external galaxies. An additional explanation is the loss of the larger aligned grains due to Radiative Torque Disruption \citep{hoan19} in very strong radiation fields, although any connection of this process with higher column depth is not clear. 

The magnetic field in the ISM is often modeled using a combination of ordered and turbulent components \citep[e.g.,][]{plan16, mivi08, jkd}. The trend of fractional polarization with column depth \citep{hild09, houd16, jone15, fiss16, plan16, plan18, jone15b} provides an indirect measurement of the effect of the turbulent component. For maximally aligned dust grains along a line of sight with a constant magnetic field direction, the fractional polarization in emission will be constant with optical depth $\tau$ in the optically thin regime. This case would correspond to a line in Figure \ref{fig:M51_IpI} with a slope of +1.0 ($\alpha = 0$). If there is a region along the line of sight with some level of variations in the magnetic field geometry, this will result in a reduced fractional polarization. Using a simple toy model, \cite{jone89} and \cite{jkd} showed that if the magnetic field direction varies completely randomly along the line of sight with a {\it{single scale length}} in optical depth $\tau$ (not physical length), then $\rm{p} \propto \tau ^{-0.5}$ (or, $\rm{I_p} \propto \tau^{+0.5}$). (See \cite{plan16, plan18} for a very similar model). In real sources, more negative slopes of $\alpha=$ -1/2 to -1 are found in many instances ranging from cold cloud cores to larger molecular cloud structures to whole galaxies  \citep[e.g.,][]{gala18, fiss16, chus19, lope19}. In more recent work employing MHD simulations, \cite{king18} and \cite{seif19} find that the ordered and random components are more complicated than modeled by \cite{jkd}. While \cite{jone15} argued that a slope of $\alpha=-1$ indicated complete loss of grain alignment due solely to loss of radiation that aligns grains by radiative torques \citep{laza07, ande15}, \cite{king19} find that including a dependency on local density for grain alignment efficiency can help explain these trends seen in large molecular clouds. 

In our large (560 pc FWHM) beam, we are averaging over many molecular clouds and associated regions of massive star formation. This complicates any effort to understand the flat slope for the upper limit in Figure \ref{fig:M51_IpI} in terms of observations and modeling for individual molecular clouds in the Milky Way. Note that the upper limit in Figure \ref{fig:M51_IpI} at larger column depths is dominated by the lower polarization in the central 3 kpc (diameter) region (see Figure \ref{fig:PitchPvsphase}). One possibility is that the field in this region has a strong component perpendicular to the plane (along our line of sight), reducing the fractional polarization. This is unlikely, given the planer field geometry seen in the central regions of edge-on spirals such as NGC 891 \citep[this paper;][]{jone97, mont14}, NGC 4565 \citep{jone97} and the Milky Way \citep[e.g.,][]{plan15}. Starburst galaxies such as M82 \citep{jone19, jone00} and NGC 4631 \citep{krau09} can show a vertical field geometry in the center, but there is no indication of a massive central starburst in M51 \citep{pine18}.  A more likely explanation is that lines of sight through higher column density paths have segments with high turbulence on smaller scale lengths ($\ll$ 560 pc) than other lower density lines of sight. In this scenario, there are segments along the line of sight that add total intensity, but add correspondingly very little polarized intensity due to turbulence in the field on scales significantly smaller than our beam (see Figure 2 in \cite{jkd}). 

The model in \cite{jkd} assumes that the optical depth scale at which magnetic field is entangled is the same through the entire volume. This may not always be true. First of all, the injection scale of the turbulence depends on the source of turbulent motions. The motions arising from large scale driving forces, whether from supernovae or magnetorotational instabilities, may have a characteristic scale comparable with the scale height of the galactic disk. The local injection of turbulence arising from local instabilities or localized energy injection sources, whatever they are, can have significantly smaller scales. These significantly smaller scales form the random component that would decrease the fractional polarization compared to the simple model. 

We also point out another important effect that affects the polarization. Even if the turbulence injection scale stays the same, the scale at which the magnetic field experiences significant changes in geometry may vary due to variations in the turbulence injection velocity. To understand this, one should recall the properties of MHD turbulence \citep[e.g.,][]{bere19}. If the injection velocity $V_L$ is larger than the Alfven velocity $V_A$, the turbulence is superAlfvenic. Magnetic forces at the injection scales are too weak to affect the motion of at large scales and at such scales the turbulence follows the usual Kolmogorov isotropic cascade with hydrodynamic motions freely moving and bending magnetic fields around. However at the scale $l_A=LM_A^{-3}$, where $L$ is the turbulence injection scale and $M_A=V_L/V_A$, the turbulence transfers to the MHD regime with the magnetic field becoming dynamically important \citep{laza06}. The scale $l_A$ is the scale of the entanglement of magnetic field. This scale determines the random walk effects on the polarization in the \cite{jkd} model. Evidently, $l_A$ varies with the media magnetization and the injection velocity. These parameters change through the galaxy and this can affect the observed fractional polarization at high column depths. \footnote{In the presence of turbulent dynamo one might expect that $I_A$ eventually reaches $L$. However, the non-linear turbulent dynamo is rather inefficient \citep{xu16} and therefore the temporal variations in the energy injection and in Alfven speed are expected to induce significant variations of $l_A$.} To explore the nature of the turbulent component further, we next compare the radio synchrotron polarimetry with our FIR polarimetry.

\subsection{Radio Comparison}\label{sec:Radio}

\begin{figure}
    \includegraphics[width=\columnwidth]{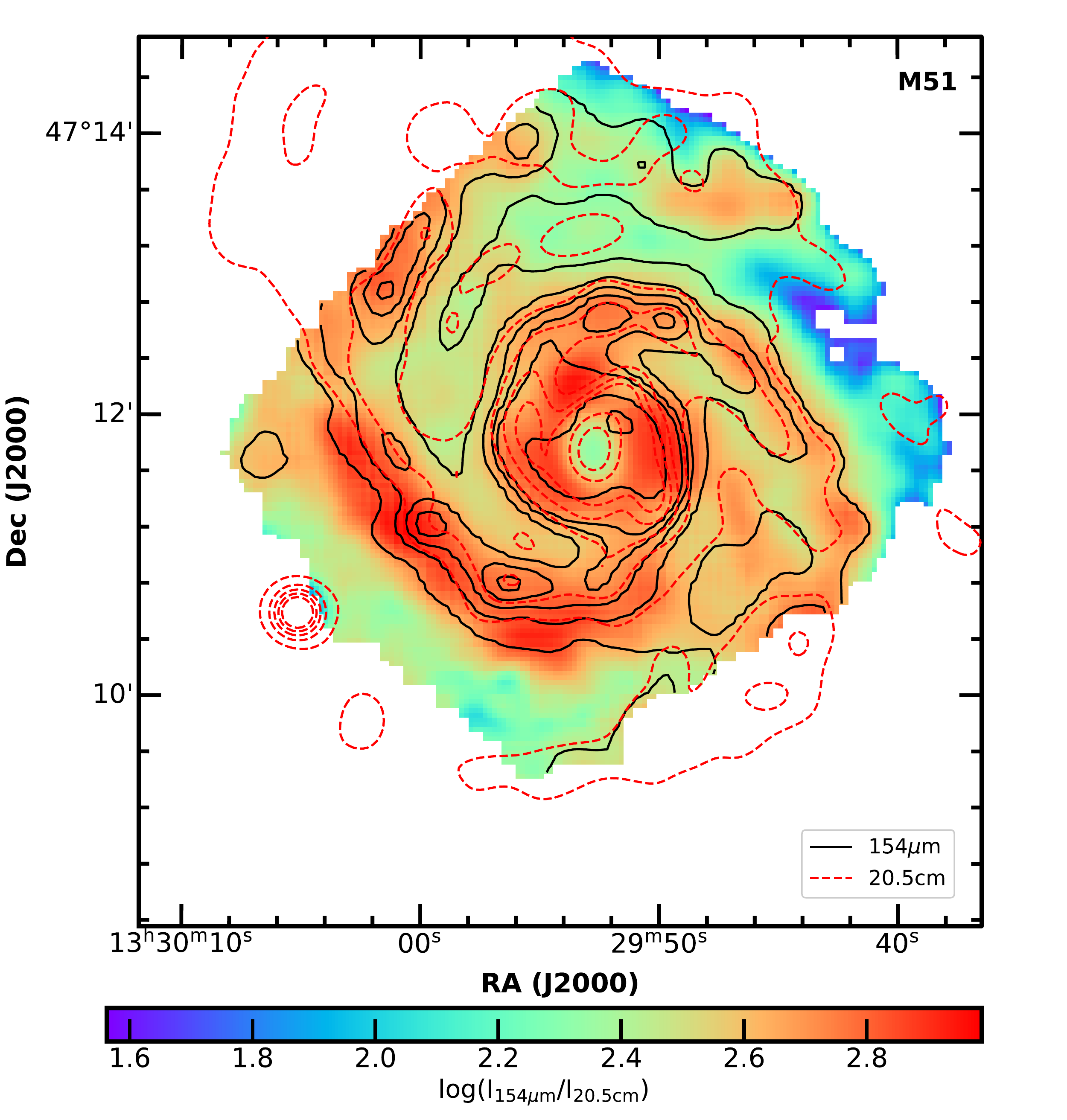}
    \caption{\label{fig:20cm_comp} The ratio of the total intensity at $154~\micron$ to that at 20.5 cm. Color represents the ratio on a logarithmic scale, $\log(\rm{I}_{154~\mu\rm{m}}/\rm{I_{20.5~cm}})$. The black contours indicate 100, 200, 300, 400, 500, 1000, and 1500 $\rm{MJy~sr^{-1}}$ at $154~\micron$ and the red contours 0.3, 0.6, 0.9, 1.2, 1.5, 3.0, and 4.5 $\rm{MJy~sr^{-1}}$ at 20.5 cm.}
\end{figure}

\begin{figure}
    \includegraphics[width=\columnwidth]{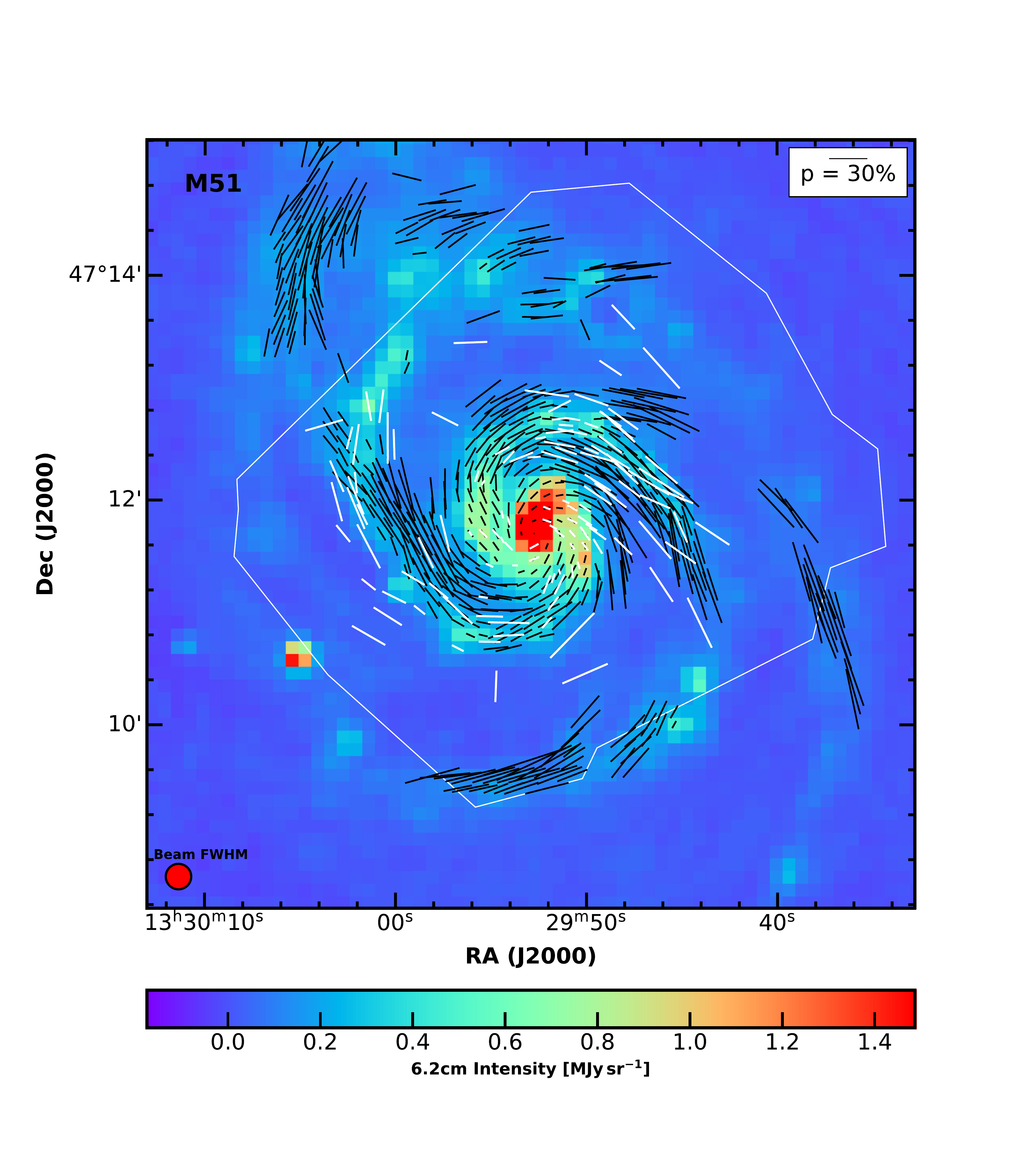}
    \caption{\label{fig:Map6cm} Fractional polarization vector maps of M51 at a wavelength of $154~\micron$ (white) and 6.2 cm (black). The colors show the intensity at 6.2 cm convolved to our beam at $154~\micron$. The scale bar for fractional polarization refers to the 6.2 \rm{cm} data only. The lengths of vectors at 154 $\mu$m are the same as those in Fig. \ref{fig:Map_final}. The thin white line roughly outlines the observed region at $154~\micron$.}
\end{figure}

\begin{figure}
    \includegraphics[width=\columnwidth]{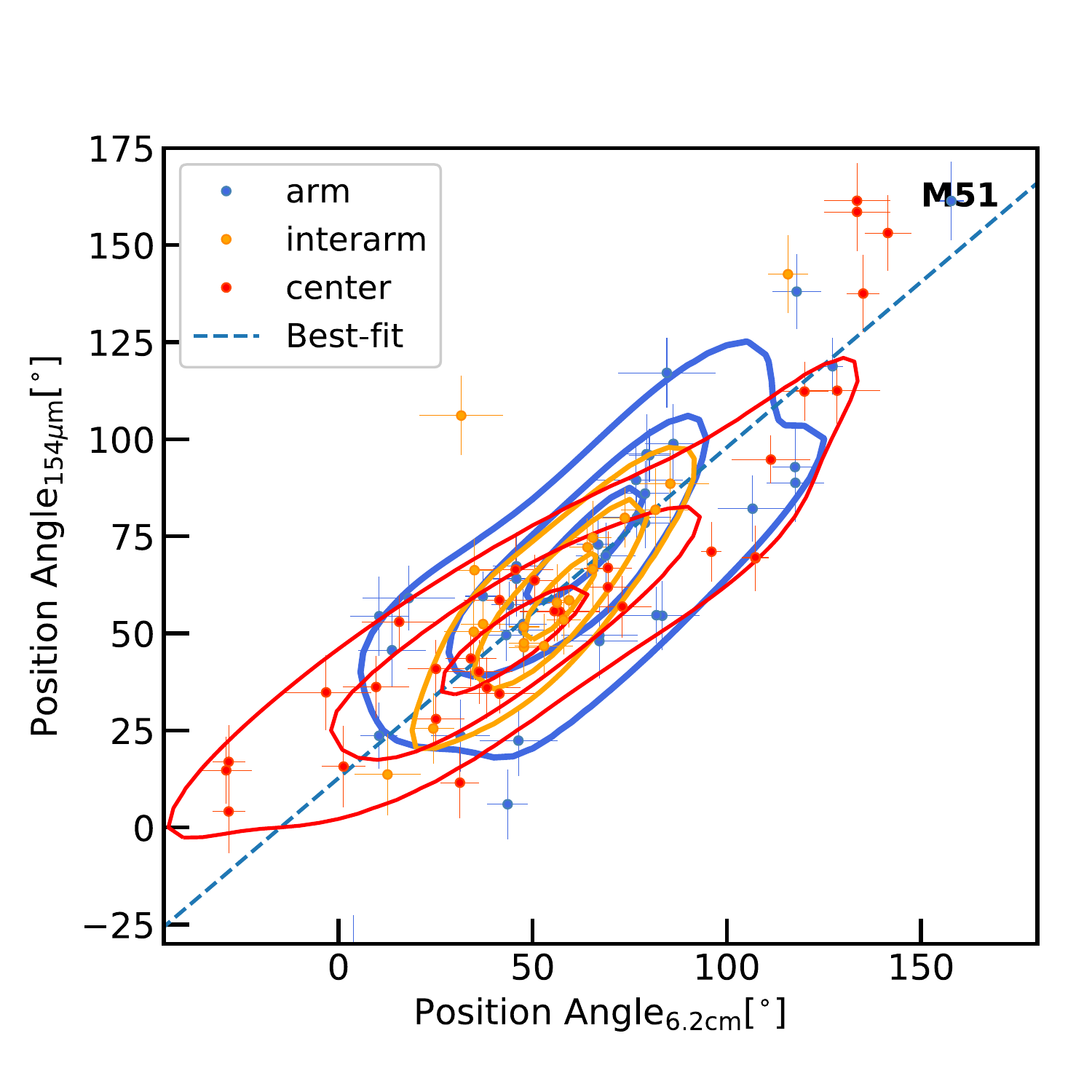}
    \caption{\label{fig:PA154_PA6} Plot of the $154~\micron$ position angle against the 6.2 cm position angle. $180\degr$ has been added to some position angles to account for the ambiguity at $0\degr$ and $180\degr$. The Pearson correlation coefficient for each region is higher than 0.75 and the p-values are smaller than $10^{-4}$. The ODR best fit line weighted by the squares of errors to all the data has a slope of 0.85 $\pm 0.12$ at the $1-\sigma$ confidence interval. The contours show the probability density of 0.3, 0.6, and 0.9 estimated by Gaussian kernel density estimation (KDE) using \texttt{scipy.stats.gaussian\_kde} module. KDE is a way to estimate the probability density function by putting a kernel on each data point, and we used Scott's Rule to determine the width of a Gaussian kernel.}
\end{figure}

\begin{figure}
    \includegraphics[width=\columnwidth]{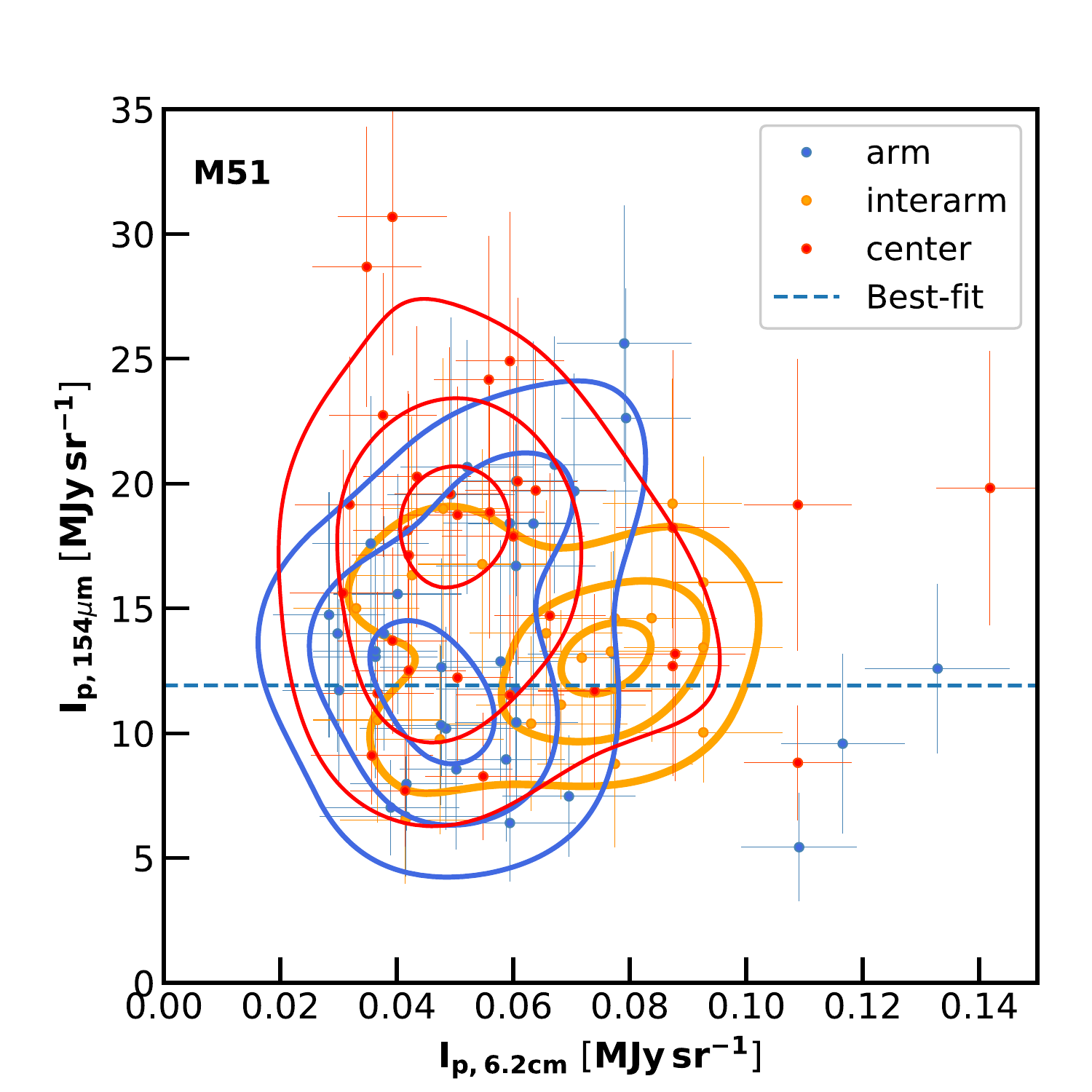}
    \caption{\label{fig:PI154_PI6} Plot of the polarized intensity at $154~\micron$ against the polarized intensity at 6.2 cm. The colors of dots indicate the different regions of arm (blue), inter-arm (orange), and center (red). The symbols and contours are the same as in Figure \ref{fig:PA154_PA6}. The Pearson correlation coefficients and p-values for the arm, inter-arm, and center are [0.014, 0.94], [0.1, 0.66], and [0.11, 0.56] respectively, indicating no correlation.}
\end{figure}

\begin{figure}
    \includegraphics[width=\columnwidth]{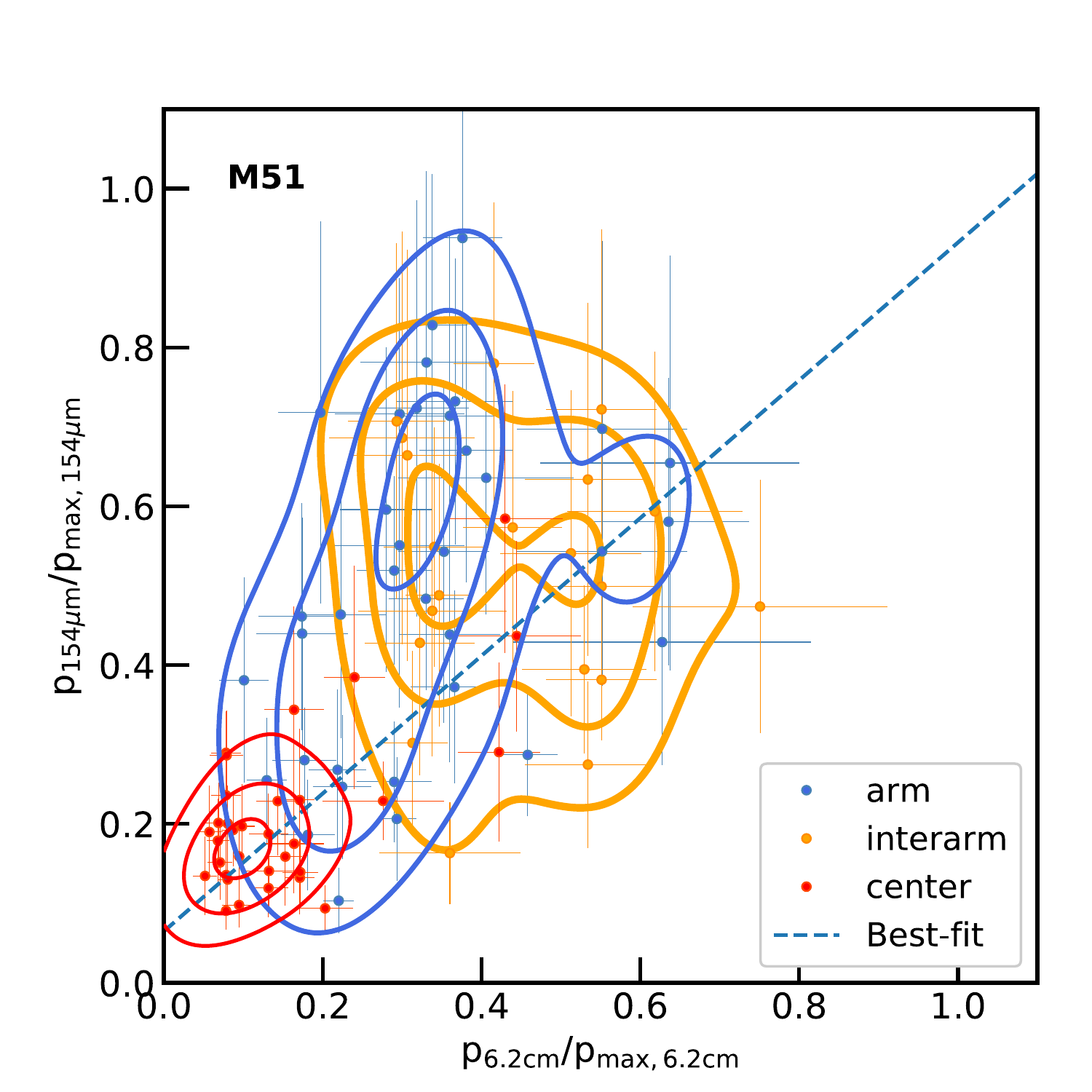}
    \caption{\label{fig:P154_P6} Plot of the normalized fractional polarization at $154 ~\micron$ against the normalized fractional polarization at 6.2 cm. The normalization factor was $9\%$ at $154~\micron$ and $70\%$ at 6.2 cm (see text). The symbols and contours are the same as in Figure \ref{fig:PA154_PA6}. The Pearson correlation coefficients and p-values for the arm, inter-arm, and center are [0.38, 0.02], [-0.06, 0.82], and [0.68, $10^{-5}$] respectively. The correlation coefficient for the entire data set is 0.61 with a p value of $10^{-9}$. The slope of the best fit line to all the data is $0.87\pm0.22$.}
\end{figure}

The magnetic field geometry of M51 seen in synchrotron polarimetry has been also been extensively studied \citep{Beck87, flet11}. We can compare the FIR emission with the synchrotron radiation at 20.5 cm and 6.2 cm using the data from \cite{flet11}, which we obtained from ATLAS OF GALAXIES at Max Planck Institute for Radio Astronomy \footnote{https://www.mpifr-bonn.mpg.de/atlasmag}. We rotated the 6.2 cm radio vector position angles by $90\degr$ to obtain the inferred magnetic field direction and made no correction for Faraday rotation (\cite{flet11} found no statistically significant difference in fractional polarization between 3.6 cm and 6.2 cm wavelengths). The beam sizes at 20.5 cm and 6.2 cm are $15''$ and $8''$ \citep{flet11}, while our beam size at $154~\micron$ is $14''$. First, in Figure \ref{fig:20cm_comp}, we compare the total intensity at $154~\micron$ and at 20.5 cm, which has a similar beam size to that at $154~\micron$. We have convolved the $154~\micron$ beam to the slightly larger beam at 20.5 cm assuming a Gaussian form for the beam shape. To be conservative in our comparison, we use only regions where all the pixels in the $154~\mu m$ image have $\rm{I / I_{err} > 5}$. In Figure \ref{fig:20cm_comp} we show the color coded intensity ratio on a logarithmic scale, $\log (\rm{I}_{154~\mu \rm{m}} / \rm{I_{20.5 cm}})$ along with the intensity contours at $154~\micron$ and 20.5 cm. 

Overall, the synchrotron emission and the FIR emission closely follow the grand design spiral pattern seen at other wavelengths. The arms are brighter than the inter-arm region at both wavelengths. However, the $154~\micron$ emission shows greater contrast between the arm and inter-arm regions compared to the 20.5 cm emission, in many locations by up to a factor of 3 greater contrast. This contrast ratio is highest in the arm to the southeast of the center, and in the arms near (but not directly at) the center of the galaxy. \cite{basu12} compared Spitzer $70~\micron$ with 20 and 90 cm radio fluxes for four galaxies and found a greater FIR/radio flux ratio in the arms compared to the inter-arm region using 90 cm radio fluxes, but not for 20 cm fluxes. Based on our $154~\micron$ fluxes and the 20.5 cm data of M51, the FIR and radio measurements are not sampling volumes along the line of sight in the same way. 

To first order, the dependence of synchrotron emission on cosmic ray electron density and magnetic field strength is $\rm{I_{syn}}\propto n_{ce}B^2$ \citep[e.g.,][]{jone74}, where $\rm{I_{syn}}$ is the synchrotron intensity and $\rm{n_{ce}}$ is the cosmic ray electron density. \cite{crut12} finds that the line of sight component (only) of the magnetic field strength (typically $2-10~\mu\rm{G}$) in the diffuse ISM of the Milky Way shows no clear trend with hydrogen density up to $n_H \sim 300~\rm{cm^{-3}}$, a density typical for photo dissociation regions and the outer edges of molecular clouds \citep{holl99}. At even higher densities the field strength increases with density as $\rm{B}\propto n_H^k$ with the exponent k between 2/3 and 1/2 \cite[e.g.][]{trit15, jian20}, but these regions occupy a small fraction of the total volume of the ISM \citep{holl99}. We interpret our results as due to the synchrotron emission in M51 arising mostly in the more diffuse ISM, with denser regions contributing a smaller fraction. Assuming equipartition between the cosmic ray energy density and the magnetic field energy density, \cite{flet11} find a moderately uniform magnetic field strength of $20-25~\mu\rm{G}$ in the arm and $15-20~\mu\rm{G}$ in the inter-arm regions of M51, suggesting the synchrotron emission is more dependent on $\rm{n_{ce}}$  than magnetic field strength in those regions. In the denser star forming regions located in the spiral arms, the ratio of FIR to radio intensity must be dominated by emission from warm dust in a volume that does not contribute as much proportionally to the total synchrotron emission as it does to the FIR emission. Note that the very center of M51 has a synchrotron emission peak \citep{quer16} due to a Seyfert 2 nucleus \citep{Ho97} emitting a relatively low luminosity of $L_{bol} \sim 10^{44} erg~s^{-1}$ \citep{Woo02}, but the FIR emission peaks outside this region in the inner spiral arms (see Figure \ref{fig:M51_spiral}), and the AGN contributes very little to the FIR flux.

For comparison of the radio and FIR polarization, we used the observations at 6.2 cm instead of 20.5 cm because depolarization in the beam by differential Faraday rotation is less \citep{flet11}. We first convolved the 6.2 cm I, Q and U maps to a $14\arcsec$ beam. We used the rms fluctuations in the convolved Q and U maps well off the galaxy to estimate the error in Q and U. Assuming these errors, the fractional polarization could then be computed and debiased in the same manner as our FIR polarimetry ($\rm{p_{debiased}} / \rm{p_{err}} > 3$), except no cut was made in the synchrotron total intensity. In Figure \ref{fig:Map6cm} we plot the resulting 6.2 cm radio and FIR polarization vectors overlayed on a map indicating radio intensity. The polarization vectors at both wavelengths clearly delineate the grand design spiral. There is good agreement in position angle at most locations where there is significant overlap, with one exception.  At 13$^{\rm{h}}$ 30$^{\rm{m}}$ 02$^{\rm{s}}$ +47$\degr$ 12$\arcmin$ 30$\arcsec$ the 6.2 cm vectors angle away from the arm along the bridge of emission connecting to M51b, but the FIR vectors continue to follow the spiral pattern.

The polarization position angles are compared quantitatively in Figure \ref{fig:PA154_PA6}, and show a strong overall correlation between the radio and FIR polarization vectors. Even though the emission mechanisms are completely different, and the ISM in the respective beams is being sampled differently, we find that the inferred magnetic field geometry is essentially the same in a global sense. In other words, the  FIR polarization position angle weighted by dust emission (at varying temperatures) integrated along and across the line of sight is very similar to the synchrotron position angle weighted by cosmic ray density and field strength (squared), integrated along the \textbf{\textit{same}} paths in most locations.

Our goal in this section is to investigate whether the synchrotron observations can shed light on the underlying cause of the strong decline in fractional polarization with intensity found at FIR wavelengths. For example, consider the hypothesis that there are segments across the beam and along a line of sight associated with dense gas and dust that have field geometries highly disordered in our beam relative to the larger scale field, adding significant FIR total intensity but very little polarized intensity. In lower column depth lines of sight, these segments (perhaps giant molecular clouds) may be absent or relatively rare, making proportionally less of a contribution to the total FIR intensity, and have less effect on the fractional polarization. Since the synchrotron polarimetry is sampling the same line of sight differently, these segments may contribute differently to the polarized synchrotron emission. 

We compare the polarized intensity between the FIR and the radio in Figure \ref{fig:PI154_PI6} and the fractional polarization in Figure \ref{fig:P154_P6}. Although this may seem redundant, there are important differences between the polarized intensity and the fractional polarization. In the diffuse ISM there is no clear dependence of dust grain alignment on magnetic field strength \citep{plan15, jone89, jone15b}. Thus, in the FIR neither polarized intensity nor fractional polarization are dependent on magnetic field strength, but they are strongly dependent on the magnetic field geometry \citep{plan18, plan16, jkd}. For synchrotron emission, the polarized intensity is dependent on magnetic field strength and the magnetic field geometry, but the fractional polarization is dependent only on the field geometry, as is the case in the FIR. Thus, we should expect no correlation between polarized intensity at the two wavelengths, but there should be a correlation between their fractional polarization if they are indeed sampling the same net magnetic field geometry.

In Figure \ref{fig:PI154_PI6}, there is $\bf{no}$ correlation seen between the polarized intensity at FIR and 6.2 cm wavelengths for the higher surface brightness central region (red contours), the arm region (blue contours), or the inter-arm region (orange contours). For fractional polarization (Figure \ref{fig:P154_P6}), we have normalized both the FIR and 6.2 cm polarization with respect to their maximum expected values. We used p$_{\rm{max}}$ = 70\% at 6.2 cm based on computational results in \cite{jone77}.   There is a modest correlation for the entire data set, with the greatest correlation in the center region. Note again that the central region has very weak fractional polarization at both wavelengths. 

For the arms (see Figure \ref{fig:PitchPvsphase}), we do not see a significant difference in fractional polarization for our FIR observations when compared to the inter-arm region. At radio wavelengths, \cite{flet11} found that the inter-arm region has a greater fractional polarization than the arms (see their Table 2), which they attribute to a more ordered field in the inter-arm region. This difference between FIR and radio observations suggests variations in the magnetic field geometry are similar between the arm and inter-arm regions as sampled by FIR polarimetry, but that the greater column depth in the arms may have caused enough Faraday depolarization across the beam to further reduce the fractional polarization at 6.2 cm. Finally, the high surface brightness central region shows very weak fractional polarization at both wavelengths. Here the radio and FIR beams must sample a more complex magnetic field geometry with highly turbulent segments across the beam and along individual lines of sight within the beam. This more complex magnetic field geometry reduces the net fractional polarization at both FIR and radio wavelengths with, perhaps, added Faraday depolarization in the beam at 6.2 cm. Polarized emission in this region is sampled differently at the two wavelength regimes, hence producing uncorrelated polarized intensities. Yet the net position angles strongly agree, the fractional polarizations are moderately correlated, and both techniques yield the same net magnetic field geometry in the beam. We will explore this interpretation more carefully in a later paper.

\section{NGC 891}

\subsection{Introduction}

At a distance of 8.4 Mpc \citep{tonr01}, NGC 891 presents an interesting case for an edge-on galaxy that is a late type spiral with similar mass and  size compared to the Milky Way \citep{kara04}. Like the Milky Way, NIR polarimetry of NGC 891 reveals a general pattern of a magnetic field lying mostly in the plane \citep{jone97, mont14}. Radio synchrotron observations are also consistent with this general field geometry, but extend well out of the disk into the halo \citep{krau09, suku91}. According to models by \cite{wood97}, highly polarized scattered light may be a contaminant affecting the optical and NIR polarization in edge-on systems producing polarization null points at locations along the disk, well away from the nucleus. \cite{mont14} do not find evidence for the predicted null points along the disk, but do find null points at other locations that they associate with an embedded spiral arm along the line of sight. Optical polarimetry \citep{scar96} revealed (unexpected) polarization mostly vertical to the plane, with only a few locations in the NE showing polarization parallel to the disk. The optical polarimetry was attributed to vertical magnetic fields, but \cite{mont14} argued that the optical polarimetry was contaminated by scattered light. Scattering in the halo of light from stars in the disk and the bulge, as modeled by \cite{wood97} and \cite{seon18}, may be a more likely explanation for the optical polarization.  Note that the NIR and FIR polarimetry penetrate much deeper into the disk than is possible at optical wavelengths. 

\subsection{The Planar Field Geometry}

Our $154~\micron$ polarimetry of NGC 891 is shown in Figure \ref{fig:NGC891_Map_final} where the colors and symbols are the same as described for M51. To show the magnetic field geometry more clearly, we set the fractional polarization to a constant value in Figure \ref{fig:NGC891_Map_final_constlength}. Along the center of the edge-on disk, the vectors align very close to the plane of the disk everywhere except in the extreme NE. There, a few vectors are perpendicular to the disk, suggesting a vertical magnetic field, which will be discussed below. Clearly evident in both the NIR polarimetry \citep{jone97, mont14} and the radio synchrotron polarimetry \citep{krau09, suku91} is an $\sim 15\degr$ tilt for many of the polarization vectors relative to the galactic plane to the NE of the nucleus. Figure 8 in \cite{mont14} best illustrates this offset, and it is not seen in the FIR vectors. 

The distribution of $\Delta \theta$ between the position angle of our rotated polarization vectors and the major axis is shown in Figure \ref{fig:NGC891_PA}. We used $21^{\circ}$ as the position angle for the major axis of the galaxy \citep{Sofue87}. In an identical manner to M51, we simulated the expected distribution under the assumption that the polarization vectors intrinsically follow the major-axis of the galaxy and only observation error causes any deviation. In Figure \ref{fig:NGC891_PA} the grey solid line shows the distribution for all the data whereas the solid, light grey bars show the  distribution only for regions with intensity higher than 1500 $\rm{MJy~sr^{-1}}$, which isolates the bright dust lane (see Figure \ref{fig:NGC891_Map_final}). When constrained to the bright dust lane, the simulated distribution and the observed distribution are very similar, with a formal p-value for this comparison is 0.97. 

Although more penetrating than optical polarimetry, NIR polarimetry at $1.65~\micron$ still experiences significant interstellar extinction in dusty, edge-on systems \citep[e.g.,][]{clem12, jone89}. In a beam containing numerous individual stars mixed in with dust, the NIR fractional polarization in extinction will saturate at ${\rm{A}}_{\rm{_V}} \sim 13$, or ${\rm{A}}_{\rm{_H}} \sim 2.5$, \citep[Fig. 4,][]{jone97}. At $154~\micron$, the disk is essentially optically thin ($\tau \sim 0.05$ for $\rm{A_V} = 100$, \cite{jone15}), thus the FIR polarimetry penetrates through the entire edge-on disk. One interpretation of our FIR polarimetry is that the NIR is sampling the magnetic field geometry on the near side of the disk, where the net field geometry shows a tilt in many locations, perhaps due to a warp in the disk \citep{oost07}. The FIR polarimetry is sampling the magnetic field geometry much deeper into the disk, where the net field geometry is very close to the plane. The radio synchrotron polarimetry at 3.6 cm from \cite{krau09} used a much larger beam of $84\arcsec$, and could be influenced by strong Faraday depolarization in the small portion of their beam that contains the disk, which has a much greater column depth than is the case for the face-on M51. Their net position angles may be sensitive only to the field geometry in the rest of the beam, also possibly influenced by the warp. Whatever the explanation, the FIR polarimetry along the disk within $2\arcmin$ of the nucleus clearly indicates that the magnetic field direction deep inside NGC 891 lies very close to the galactic plane.

There are two regions of enhanced intensity in the disk about $1\arcmin$ on either side of the nucleus, designated by colored outlines in Figure \ref{fig:NGC891_Map_final}. These locations also correspond to intensity enhancements seen in a radio map of the galaxy made by combining LOFAR and VLA observations \citep{mulc18}, and in PACS $70~\micron$ observations as well \citep{bocc16}. Those studies attribute such enhancements to the presence of spiral arms and the enhanced star formation associated with them, but do not present a model of the emission from the disk. These features are $3-4$ kpc from the center, not untypical for spiral arms. For example, rotate M51 about a N-S axis to create an edge-on spiral, and there would be enhancements in FIR emission on either side from the center at this distance. The polarization is very low in the southern region, at the limits of our detection. The polarization is also quite low in the northern bright spot. As with M51 and discussed below for NGC 891, the fractional polarization is anti-correlated with intensity, so this may not be unexpected, but the polarization in the southern spot in particular is exceptionally low. \cite{mont14} also found regions along the disk where the NIR polarimetry was very low. They suggested the observer was looking down along a spiral arm, where the magnetic field is largely $along$ (parallel to) the line of sight, which results in much lower polarization \citep[e.g.,][]{jowh15}. This could be the explanation for the very low polarization in our two bright spots, and could also explain the origin of the enhancement in intensity, since a line of sight down a spiral arm will pass through more star forming regions. However, the regions of low polarization seen at NIR wavelengths and FIR wavelengths are not coincident, rather the NIR null points are located further out from the center of the galaxy. Given the greater penetrating power of FIR observations, it is possible we are viewing more deeply embedded spiral features than is accessible by NIR polarimetry, which is more sensitive to the front side of the disk. 

\subsection{Vertical Fields}

Dust in emission is detected above and below the disk of NGC 891. At FIR wavelengths, \cite{bocc16} find a thick disk component to the dust emission with a scale height of $\sim 1.5$ kpc ($36\arcsec$).  At NIR wavelengths, \cite{aoki91} measure a scale height of 350pc $(8.6\arcsec)$ for the stellar component, significantly smaller than the dust scale height. There are a handful of vectors in Figure \ref{fig:NGC891_Map_final} that lie off the bright disk in the halo of NGC 891. Five of these vectors are consistent with a vertical magnetic field geometry, in strong contrast to the disk. At optical wavelengths, \cite{howk97} imaged vertical fingers of dust that stretch up to 1.5 kpc off the plane, also suggestive of a vertical field extending into the halo. Optical polarimetry of the NE portion of the disk \citep{scar96} has a few vectors parallel to the plane, but the majority are perpendicular to the plane. Although the optical polarimetry was interpreted as evidence for vertical magnetic fields by \cite{scar96}, the NIR polarimetry from \cite{mont14} and modeling by \cite{wood97} and \cite{seon18} indicate that scattering of light originating from the central region can be a major effect. Without significant dust to shine through (causing interstellar extinction), it is difficult to produce measurable interstellar polarization in $extinction$ \citep{jowh15}. 


The optical polarization vectors in \cite{scar96} are typically 1--2 \% in magnitude $\sim 20\arcsec$ off the plane using a $12\arcsec$ beam. Based on our $154~\micron$ contours, this corresponds to about $400~\rm{MJy~sr^{-1}}$, or $\rm{A_V} \sim 0.4$. The historically used empirical maximum for interstellar polarization in extinction at V is $\rm{p(\%)}=3\rm{A_V}$ \citep{serk75}, but recent work shows this can be as high as $\rm{p(\%)}=5\rm{A_V}$ for low density lines of sight out of the Galactic Plane \citep{pano19}. For an optimum geometry of a screen of dust with a uniform magnetic field geometry entirely in front of the stars in the halo, a maximum fractional polarization of $\sim 2\%$ would be expected. For a mix of dust and stars along the line of sight and turbulence in the magnetic field, the expected fractional polarization would be even less. Although \cite{howk97} estimated $\rm{A_V} \sim 1$ within some of the vertical filaments, which are only $2-3\arcsec$ wide, considerable unpolarized starlight emerging between the filaments would be contributing as well. At optical wavelengths it is not clear there is enough extraplanar dust to shine through to cause significant polarization in extinction $\sim 20\arcsec$ off the disk, but plenty of dust to scatter light (a mean $\tau_{sc}\sim 0.3$ at V) from stars in the disk and bulge. As with M51, the striking similarity between the optical polarimetry vectors and our FIR vectors can not be denied, and remains a mystery when the non-detection at NIR wavelengths is considered.

Polarimetry at FIR wavelengths is measuring the $emission$ from warm dust, and generally the fractional polarization is observed to be highest at low FIR optical depths \citep{chus19, plan15, fiss16}, but there must be enough warm dust in the beam to produce a measurable signal. For our observations of NGC 891, a vertical scale height of 1.5 kpc corresponds to $36\arcsec$, or 2.7 beamwidths for our $154~\micron$ observations. The surface brightness at this vertical distance for most of the disk is $\sim 100~\rm{MJy~sr^{-1}}$ ($\rm{A_V}\sim 0.1$), which is near the limit of our detectability of statistically significant fractional polarization. At 1.5 beams ($20\arcsec$) off the plane, the surface brightness ranges from $300~\rm{MJy~sr^{-1}}$ to $500~\rm{MJy~sr^{-1}}$, a range in which 5\% polarization is easily detectable. Note, if NGC 891 were face-on, this halo dust emission would contribute very little to the total flux in our beam compared to the disk. 

We draw the tentative conclusion that the several $154~\micron$ vectors in the halo that are perpendicular to the disk are indicative of a vertical magnetic field geometry in the halo of NGC 891. No evidence for vertical fields was found in radio observations by \cite{krau09}, but they had a very large $84\arcsec$ beam. Using a $20\arcsec$ beam, \cite{suku91}  find hints of a vertical field on the eastern side of the southwest extension of the disk, just east of the region outlined in green in Figure \ref{fig:NGC891_Map_final}, where we suggest we are looking down a spiral arm. \cite{mora19} made radio observations of NGC 4631, an edge-on galaxy with an even more extended halo than NGC 891, using a $7\arcsec$ beam. They find the magnetic field in the halo is characterized by strong vertical components. Examination of the Faraday depth pattern in the halo of NGC 4631 indicated large-scale field reversals in part of the halo, suggesting giant magnetic ropes, oriented perpendicular to the disk, but with alternating field directions. Our FIR polarimetry, which is not affected by Faraday rotation, cannot distinguish field reversals (since the grain alignment is the same), and would reveal only the coherent, vertical geometry, such as we see in our observations in the halo of NGC 891. \cite{bran20} present numerical results of mean-field dynamo model calculations for NGC 891 as a representative case for edge-on disk systems, but our observations do not have enough vectors for a detailed comparison.

\subsection{Polarization -- Intensity Relation}

\begin{figure}
    \includegraphics[width=\columnwidth]{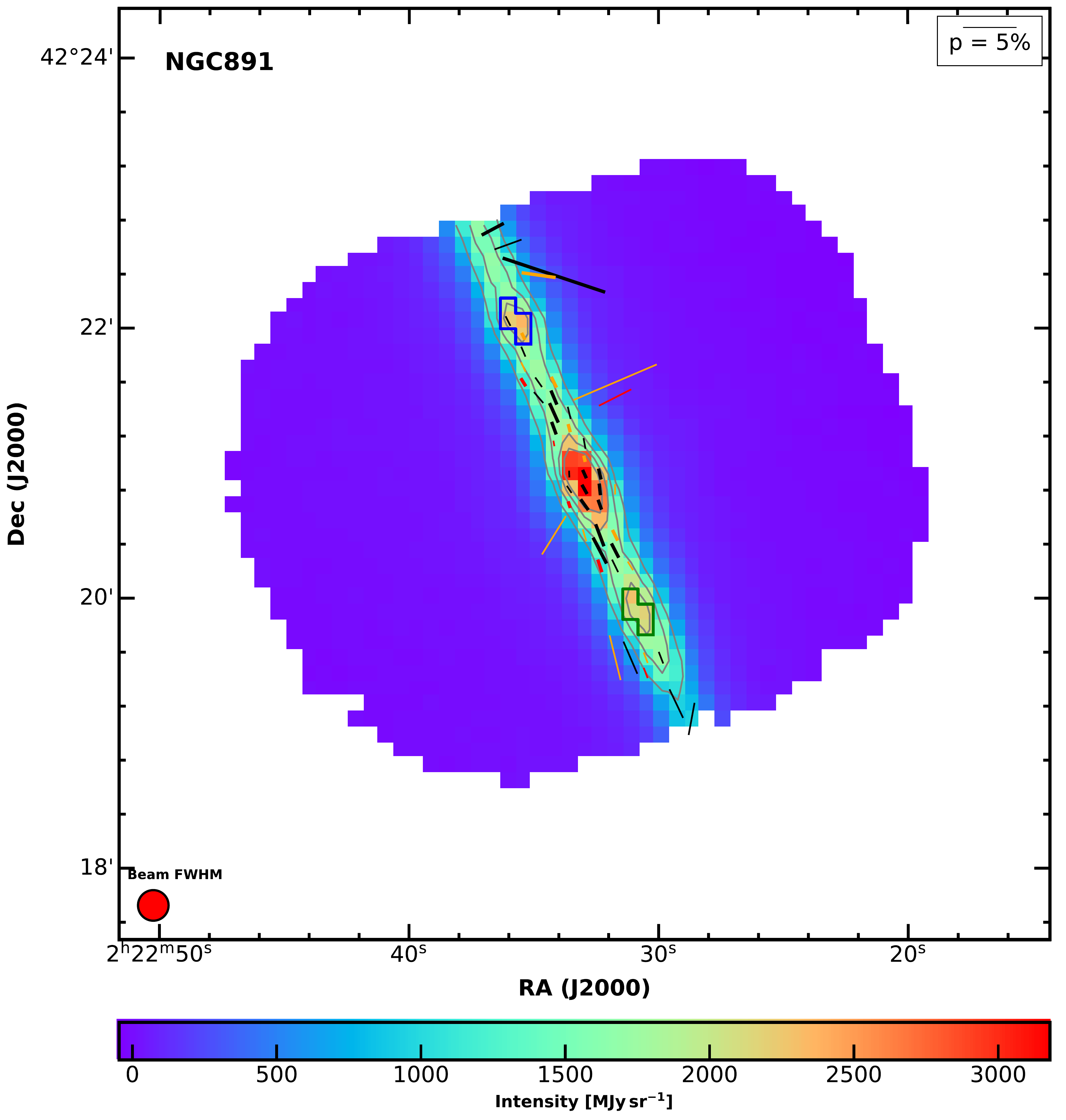}
    \caption{\label{fig:NGC891_Map_final} Polarization vector map of NGC891 at a wavelength of $154~\micron$, in which the E vectors are rotated $90^{\circ}$ to represent the inferred magnetic field direction. Data points using a square $6.8 \arcsec \times 6.8 \arcsec$ `half' beam are plotted in black. Data points using a $13.6 \arcsec \times 13.6 \arcsec$ `full' beam are plotted in orange, and red vectors are computed using a $27.2 \arcsec \times 27.2\arcsec$ square beam. The red disk in the lower left corner indicates the FWHM footprint of the HAWC+ beam on the sky at $154~\micron$. Vectors with S/N $\geq 3:1$ have thick lines and vectors with S/N from 2.5:1 to 3:1 have thin lines. The color map represents the $154~\micron$ continuum intensity and grey contours show 1000, 1500, 2000, 2500 $\rm{MJy~sr^{-1}}$. Two regions discussed in the text are outlined by blue and green boxes.}
\end{figure}

\begin{figure}
    \includegraphics[width=\columnwidth]{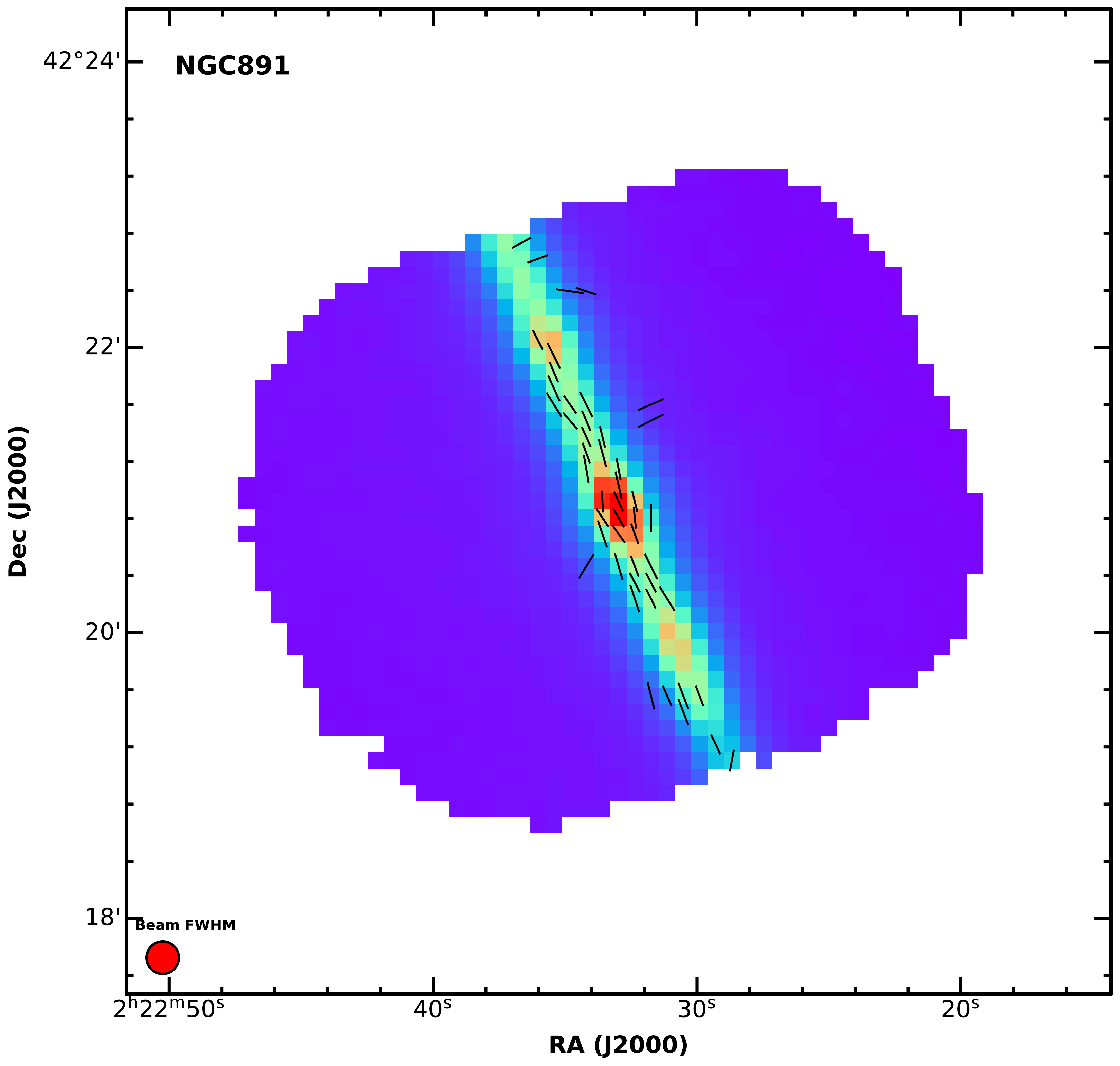}
    \caption{\label{fig:NGC891_Map_final_constlength} Same as Figure \ref{fig:NGC891_Map_final}, except all the polarization vectors have been set to the same length to better illustrate the position angles.}
\end{figure}

\begin{figure}
    \centering
    \includegraphics[width=.5\columnwidth]{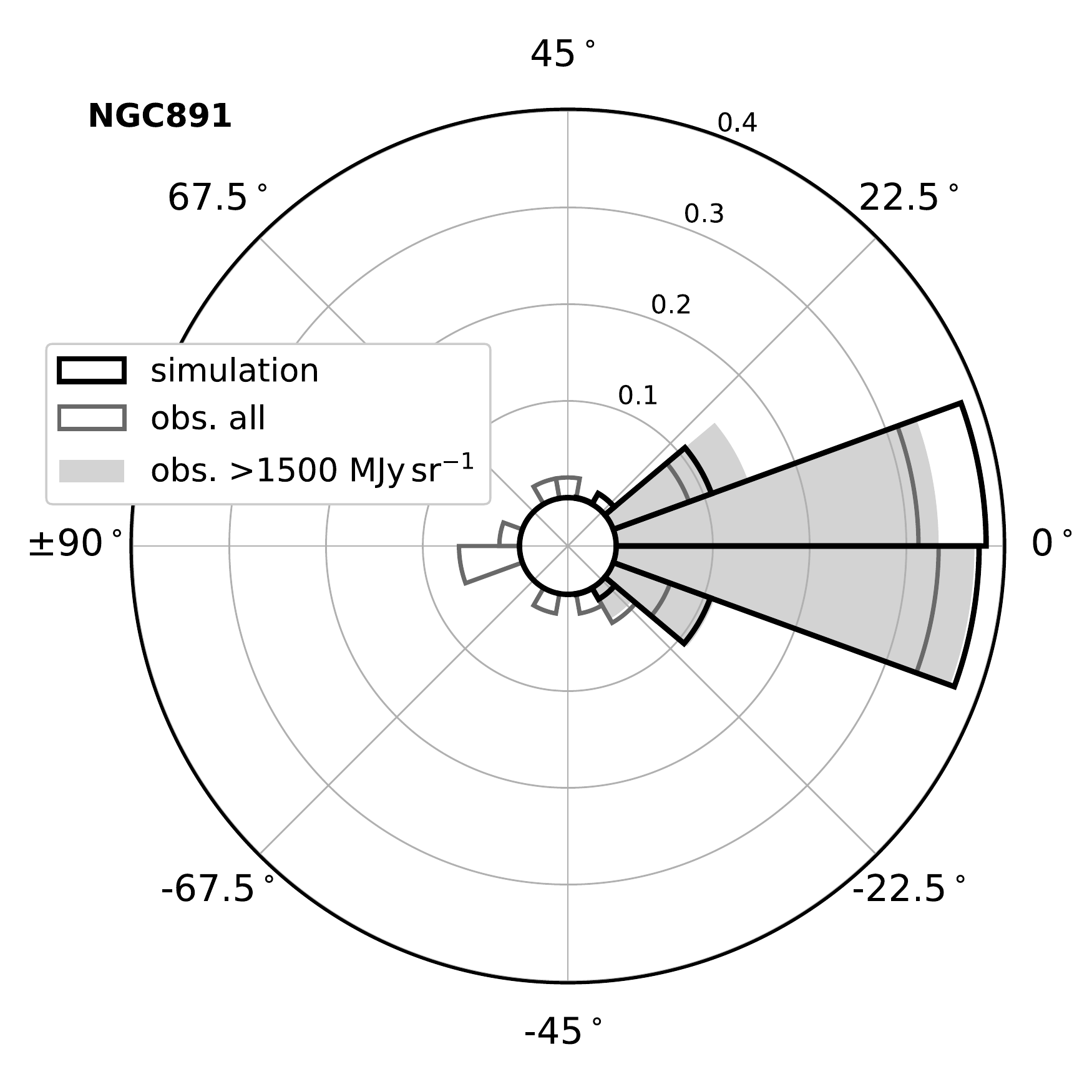}
    \caption{\label{fig:NGC891_PA} Distribution of $\Delta \theta$ between the position angle of our polarization vectors and the major-axis the galaxy. A positive value means counter-clockwise rotation from the major-axis. The Grey solid line shows the distribution of all data and the grey shaded region that of the data only in the region with intensity higher than 1500 $\rm{MJy~sr^{-1}}$. The black solid line indicates a simulation made under the assumption that the polarization vectors follow the major-axis of the galaxy and only errors in the data contribute to the dispersion.}
\end{figure}

\begin{figure}
    \includegraphics[width=\columnwidth]{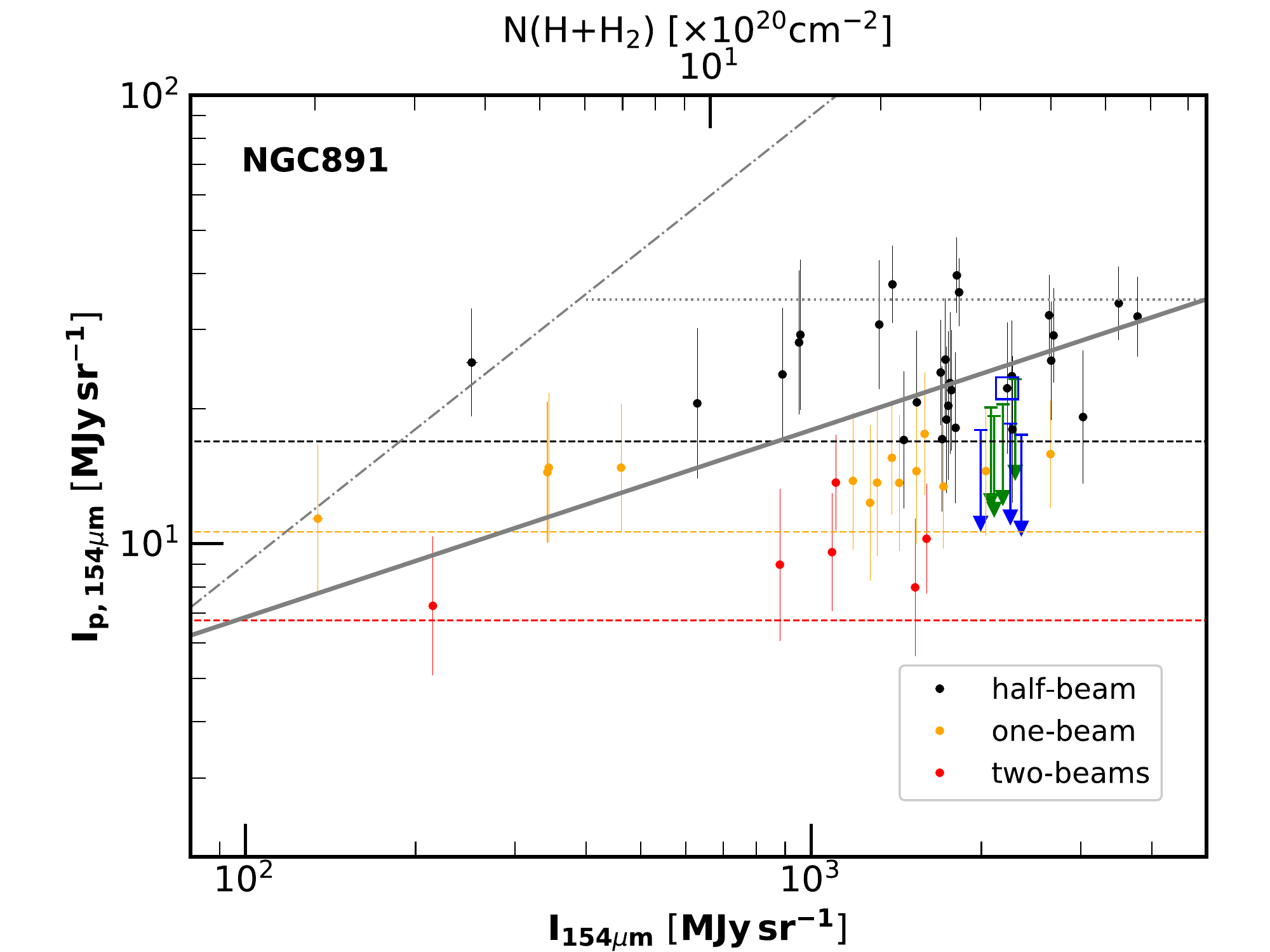}
    \caption{\label{fig:NGC891_IpI} Plot of the polarized intensity against the intensity at $154~\micron$. The vectors shown in Figure \ref{fig:NGC891_Map_final} were used.  A grey solid line is a fit to the data, where $\log \rm{I}_{\rm{p}154~\micron} = 0.42 \log \rm{I}_{154~\micron}$. All other lines are the same as in Figure \ref{fig:M51_IpI}. The green and blue upper limits and boxed blue points are described in Section 4.4 of the text.}
\end{figure}

Figure \ref{fig:NGC891_IpI} plots the polarized intensity  against the intensity and column depth for NGC 891. Other than using a temperature of 24 K for the dust \citep{Hughes14}, the procedure for calculating the column depth from the surface brightness at $154~\micron$ is the same as for M51. NGC 891 shows a clear trend in $\rm{I_p}$ vs. I, with a similar slope to that found for M51, and shows evidence for a horizontal upper limit as well. However, unlike M51, the decrease in polarization in the bulge is not quite as strong, and more of the very low fractional polarization values are located in the disk away from the nucleus. Also unlike M51, the data at lower column depth in either the disk or the halo generally lie well below the upper limit of $\rm{p}=9\%$ in Figure \ref{fig:NGC891_IpI}, although this may be partially due to smaller number of vectors compared to M51. Presumably, the more complex line-of-sight magnetic field geometry through an edge-on galaxy reduces the net polarization compared to the face-on geometry for M51. Spiral structure seen edge-on can present a range of projected magnetic field directions along a line of sight, crossing nearly perpendicular to some arms, but more down along other arms in our beam.

The two regions with low polarization delineated in Figure \ref{fig:NGC891_Map_final} by green and blue outlines are shown in Figure \ref{fig:NGC891_IpI} using the same colors. These are the two regions we speculated were lines of sight down a spiral arm, reducing the fractional polarization. There is only one detection in these regions and all the rest of the data points are $3\sigma$ upper limits, indicating a low fractional polarization compared to the general trend. Until a model of the spiral structure in NGC 891 is developed, we can only identify these two locations as potential indicators of spiral features.

\section{Conclusions}

In this work we report $154~\micron$ polarimetry of the face-on galaxy M51 and the edge-on galaxy NGC 891 using HAWC+ on SOFIA with projected beam sizes of 560 and 550 pc respectively. We have drawn the following conclusions:

1. For M51, the FIR polarization vectors (rotated $90\degr$ to infer the magnetic field direction) generally follow the spiral pattern seen in other tracers. The dispersion in position angle with respect to the spiral features is greater than can be explained by observational errors alone. For the arm region, the position angles may be consistent with the spiral pattern, but uncertainties in the contribution of a random component to the magnetic field prevents us from making a more definitive statement. The central region, however, clearly shows a more open spiral pattern than seen in the CO and dust emission. 

2. Even though the FIR (warm dust) and 6.2 cm (synchrotron) emission mechanisms involve completely different physics and sample the line-of-sight differently, their polarization position angles are well correlated. The ordered field in M51 must connect regions dominating the synchrotron polarization and the FIR polarization in a simple way.

3. Both the 6.2 cm synchrotron and FIR emission show very low fractional polarization in the high surface brightness central region in M51. There is a moderate correlation in fractional polarization between the two wavelengths, yet the polarized intensity shows no correlation anywhere in the galaxy. The low polarization is likely caused by an increase in the complexity of the magnetic field and a greater contribution from more turbulent segments in the beam and down lines of sight within the beam. The lack of correlation between polarized intensity at both wavelengths indicates that the magnetic field strength, which influences the polarized intensity at 6.2 cm, but not in the FIR, is not the cause of the low fractional polarization at FIR wavelengths. Lack of grain alignment can also be ruled out. We conclude that along individual lines of sight, different segments must be contributing to the total and polarized intensity in different proportions at the two wavelengths.

4. Within the arms themselves, we find a similar fractional polarization to the inter-arm region in dust emission, unlike the synchrotron emission, which has a lower fractional polarization in the arms relative to the interarm region. This suggests the turbulent component to the magnetic field (as sampled by FIR emission) is similar to that in the inter-arm region, but that the synchrotron emission may be additionally influenced by some Faraday depolarization in the arms.

5. For NGC 891, the FIR vectors within the high surface brightness contours of the edge-on disk are tightly constrained to the plane of the disk. Dispersion in position angle about the plane can be explained by errors in the measurements alone. This result is in contrast to radio and NIR polarimetry which show a clear departure from planar at many locations along the disk. We are probably probing deeper into the disk of NGC 891 than is possible with NIR and synchrotron polarimetry, revealing a very planar magnetic field geometry in the interior of the galaxy.

6. There are two locations along the disk of NGC 891 that show very low polarization and may be locations where the line of sight is along a major spiral arm, resulting in lower fractional polarization. These two locations line up with FIR intensity contours, but do not correspond to nulls in the NIR polarimetry, thought to be due to the same cause. Likely, the NIR is sensitive to spiral features that are closer to the front side of the disk due to extinction obscuring such features deeper into the disk.

7. There is tentative evidence for the presence of vertical fields in the FIR polarimetry of NGC 891 in the halo that is not present at NIR wavelengths and is only hinted at in radio observations. At FIR wavelengths there is dust above and below the disk in emission, but this dust may not be enough to produce polarization in extinction at optical or NIR wavelengths.  

These data are the first HAWC+ observations of M51 and NGC 891 in polarimetry mode. The brighter regions within the spiral arms of M51 and the disk of NGC891 are well measured. However, the inter-arm regions in M51 and the halo of NGC 891 are less well measured, and these two regions will require deeper observations to better quantify the arm-- inter-arm comparison in M51 and the presence of vertical fields in NGC 891.

\section {acknowledgements}

We thank Larry Rudnick for many useful discussions on radio polarimetry.

Portions of this work were carried out at the Jet Propulsion Laboratory, operated by the California Institute of Technology under a contract with NASA.

The authors wish to thank Northwestern's Center for Interdisciplinary Exploration and Research in Astrophysics (CIERA) for providing technical support during the development and usage of the HAWC+ data analysis pipeline.

A.L. acknowledges support from National Science Foundation grant AST 1715754.

\end{document}